\documentclass[a4paper,twocolumn,aps,pra,showpacs]{revtex4}
\usepackage[latin9]{inputenc}
\usepackage{float}
\usepackage{amsmath}
\usepackage{amssymb}
\usepackage{graphicx}
\usepackage{esint}

\makeatletter

\@ifundefined{textcolor}{}
{%
 \definecolor{BLACK}{gray}{0}
 \definecolor{WHITE}{gray}{1}
 \definecolor{RED}{rgb}{1,0,0}
 \definecolor{GREEN}{rgb}{0,1,0}
 \definecolor{BLUE}{rgb}{0,0,1}
 \definecolor{CYAN}{cmyk}{1,0,0,0}
 \definecolor{MAGENTA}{cmyk}{0,1,0,0}
 \definecolor{YELLOW}{cmyk}{0,0,1,0}
}

\usepackage{mathrsfs}\usepackage{subfigure}\usepackage{multirow}\usepackage[title,titletoc,header]{appendix}

\makeatother
\begin{document}

 \draft
 \title{Generalized Josephson relation for conserved charges in multi-component bosons }
 \author{Yi-Cai Zhang}
 \address{Department of Physics, College of Physics and Electronic Engineering, Guangzhou University, Guangzhou
510006, China,
  }
\date{\today}

\begin{abstract}

The Josephson relation is generalized for conserved charges in multi-component bosons. With linear response theory, a formula for derivation of generalized superfluid density is given.  When there are several conserved charges, the superfluid density is generally a second order tensor in internal spin space. When the rank of Green's function matrix is one, Josephson relation is given explicitly with phase operator method.
For two-component bosons, with quantum field theory, we show a generalized Hugenholtz-Pines relation hold and existence of two gapless phonons. When the interactions are $U(2)$ invariant, we show there is a gapless quadratic dispersion excitation no matter how strong the interactions are. The corresponding generalized Josephson relation is expressed with Green's function matrix elements.
\end{abstract}

\maketitle
\section{Introduction }
Bose-Einstein condensation (BEC) and superfluidity are two closely related phenomenons in Bose quantum liquid at low temperature \cite{Tisza,BOGOLIUBOV1947}. Generally speaking, condensate density is not equal to superfluid density in interacting bosons \cite{Penrose,Pollock}.
B. D. Josephson derived a remarkable relation between the superfluid density $\rho_s$ and condensate density $n_0$ (the so-called Josephson relation) which hold no matter whether the system is at zero or finite temperature \cite{Bogoliubov61,Josephson1966,Hohenberg1965}
 \begin{align}
\rho_s=-lim_{q\rightarrow0}\frac{n_0m}{q^2 G(\textbf{q},0)},
\end{align}
 where $G(\textbf{q},0)$ is normal Green's function at zero frequency, $m$ is particle mass.
 The above Josephson relation in superfluid system can be viewed as a manifestation of Bogoliubov's ``$1/q^2$" theorem  for Green's function in spontaneous symmetry breaking system \cite{Bogoliubov,Forster}.
 It also has close connections with the absence of long ranged order (e.g., condensation) at finite temperature in one and two  dimensions \cite{Baym}.
Such relation had been applied to the Bose-Einstein condensate near  critical temperature  $T_c$ of condensation transition to get the scaling law of superfluid density \cite{Holzmann2003}.
 The relation has been re-derived  using various methods \cite { Dawson,Griffin,Holzmann} and its possible generalization in fermion BCS-BEC crossover \cite{Taylor} and disordered superfluid system \cite{Muller} has been investigated .

Multi-component Bose-Einstien condensate (spinor-BEC) has been predicted and realized in optically trapped alkali metal atomic gas \cite{Ho1998,Machida,Kawaguchi2012}, dependent on interaction and external Zeeman field, which can show various ground state phases \cite{Ueda}.
Another interesting  example of multiple component bosons is the spin-orbit coupled Bose-Einstein condensate \cite{Lin1}. It is shown that, because of breaking of Galilean invariance and presence of up branch excitation, even at zero temperature, there exists  a finite normal density. Consequently the superfluid density is suppressed significantly \cite{normaldensity, stringari2016}.
All the multiple-component bosons are expected to show superfluid behaviors at low temperature.
 Above these ground states, the collective modes are characterized by density or spin oscillations. In general, the mass current and spin current may coexist in multi-component bosons.
Naturally, one important question arises: how to generalize the definition for superfluid density in multiple component bosons. Furthermore, how to generalize the Josephson relation for  generalized spins (conserved charges) in multi-component system.

In order to resolve above questions, in this work, from linear response theory, we give a general definition of superfluid density $\rho_s$ and generalize Josephson relation for conserved charges in multi-component bosonic system.
Furthermore, we illustrate above discusses by several specific examples.
The paper is organized as follows. In Section. \textbf{II}, we give a review of the derivations of Josephson relation in usual single component bosons.  In order to generalize the Josephson relation, from the linear response theory, we give a general formula in Section. \textbf{III}.
  In Sec. \textbf{IV},  based on the general formula, we give several examples to illustrate above results.
   A summary is given in Sec. \textbf{V}.

\section{review of Josephson relation in usual single-component bosons}
In the rest of paper, we take $m=\hbar=V=1$, where $m$ is particle mass, $V$ is volume of system.
The Josephson relation gives a connection of superfluid density $\rho_s$, condensate density $n_0$ and single particle Green's function at zero frequency and low momentum limit \cite{Baym,Ueda,Holzmann}
\begin{align}
\rho_s=-lim_{q\rightarrow0}\frac{n_0}{q^2 G(\textbf{q},0)},
\end{align}
where single particle retarded Green's function is
\begin{align}
G(\textbf{q},\omega+i\eta)=\sum_n[\frac{|\langle0|\psi_\textbf{q}|n\rangle|^2}{\omega+i\eta-\omega_{n0}}-\frac{|\langle0|\psi^{\dag}_\textbf{q}|n\rangle|^2}{\omega+i\eta+\omega_{n0}}],
\end{align}
with bosonic field operator of momentum space $\psi_\textbf{q}$ and $\psi^{\dag}_\textbf{q}$. $\omega_{n0}=E_n-E_0$, $E_n$ and $|n\rangle$ are system eigenenergy and eigenstates, respectively.

Here we outline how to get above relation in usual single component bosonic system. Firstly we adopt the London's trial wave function to describe the superflow near ground state \cite{Pines}, namely,
 \begin{align}
 |\psi\rangle=e^{i\sum_i\delta\theta(\textbf{r}_i)}|\psi_0\rangle,
\end{align}
where $|\psi_0\rangle$ is ground state wave function and sum over all the particle's position. $\delta\theta$ is real function which denotes the phase variation.
With the above wave function, substituting it into the current operator, the resultant current (superflow) is given by
 \begin{align}
 \delta\textbf{j}(\textbf{r})\equiv\rho_{s}\vec{\nabla}\delta\theta(\textbf{r})=\rho_{s}\delta\textbf{v}_s,
 \label{deltaj}
\end{align}
 where we define superfluid velocity  as gradient of phase variation, i.e., $\delta\textbf{v}_s\equiv\vec{\nabla} \delta\theta(\textbf{r})$ and superfluid density $\rho_s$ as the coefficient before $\delta\textbf{v}_s$ in the current $\delta\textbf{j}(\textbf{r})$. On the other hand,
$\delta\theta(\textbf{r})$ is also phase variation of condensate order parameter $\langle\psi(\textbf{r})\rangle$ (the amplitude variation of $\langle\psi(\textbf{r})\rangle$ can be neglected near ground state)
\begin{align}
 &\delta \langle \psi\rangle\equiv \langle\psi\rangle e^{i\delta\theta(\textbf{r})}-\langle\psi\rangle\simeq i\langle\psi\rangle \delta\theta(\textbf{r}),
\label{deltaphi}
\end{align}
where $\langle\psi\rangle\equiv \langle\psi(\textbf{r})\rangle=\sqrt{n_0}$ is order parameter. We will see the connection between the above two equations (eqn.(\ref{deltaj}) and (\ref{deltaphi})) would give the Josephson relation.

In order to get the relationship between superfluid density $\rho_s$ and condensate density $n_0$, here one need add perturbation which couples to the annihilation operator $\psi$ and its adjoint $\psi^{\dag}$ \cite{Baym,Ueda}
\begin{align}
 H'\!\!&&&=\!\!\int \!\! d^3\textbf{r}[e^{i(\textbf{q}\cdot \textbf{\textbf{r}}-\omega t)+\eta t}\xi \psi^{\dag}(\textbf{r})\!\!+\!\!e^{-i(\textbf{q}\cdot \textbf{r}-\omega t)+\eta t}\xi^* \psi(\textbf{r})],\notag\\ &&&=\xi\psi^{\dag}_{\textbf{q}}e^{-i\omega t+\eta t}+\xi^*\psi_{\textbf{q}}e^{i\omega t+\eta t},
\end{align}
where $\xi$ is small complex number and we use relation $\psi(\textbf{r})=\sum_\textbf{q} \psi_\textbf{q}e^{i\textbf{q}\cdot \textbf{r}}$. Here we add a infinitesimal positive number $\eta\rightarrow0_+$ in the above exponential which corresponds choosing boundary condition that the perturbation is very slowly added to the system.

We assume initially the system is in the ground state $|0\rangle$, and then slowly turn on perturbation $H'$, the wave function can be written as
\begin{align}
 &&\psi(t)=\sum_n a_n(t)e^{-iE_nt}\psi_n
\end{align}
where $a_n(t\rightarrow -\infty)=\delta_{n,0}$ and $H\psi_n=E_n\psi_n$, $H$ is unperturbated Hamiltonian.
Using perturbation theory, we get
\begin{align}
 &&&\psi(t)\simeq \psi_0e^{-iE_0 t}+\sum_{n\neq0} a_n(t)e^{-iE_nt}\psi_n,\notag\\
 &&& a_n(t)=\frac{1}{i}\int_{-\infty}^{t} d\tau H'_{n0}(\tau)e^{i\omega_{n0}\tau}\notag\\
 &&&\!\!=\!\![\frac{\xi\langle n|\psi^{\dag}_{\textbf{q}}|0\rangle e^{-i(\omega+i\eta-\omega_{no})t}}{\omega+i\eta-\omega_{no}}-\frac{\xi^*\langle n|\psi_{\textbf{q}}|0\rangle e^{i(\omega-i\eta+\omega_{no})t}}{\omega-i\eta+\omega_{no}}],\notag\\
\end{align}
where $\omega_{n0}=E_n-E_0$.
The changes of order parameter $\langle\psi(\textbf{r})\rangle$ and current $\textbf{j}(\textbf{r})$ are
\begin{align}
 &&&\delta\langle\psi(\textbf{r})\rangle\notag\\
 &&&=\xi e^{-i(\omega+i\eta)t}[\frac{\langle0|\psi(\textbf{r})|n\rangle\langle n|\psi^{\dag}_{\textbf{q}}|0\rangle}{\omega+i\eta-\omega_{n0}}-\frac{\langle0|\psi^{\dag}_{\textbf{q}}|n\rangle\langle n|\psi(r)|0\rangle}{\omega+i\eta+\omega_{n0}}]\notag\\
 &&&+\xi^* e^{i(\omega-i\eta)t}[\frac{\langle0|\psi_{\textbf{q}}|n\rangle\langle n|\psi(\textbf{r})|0\rangle}{\omega-i\eta-\omega_{n0}}-\frac{\langle0|\psi(\textbf{r})|n\rangle\langle n|\psi_{\textbf{q}}|0\rangle}{\omega-i\eta+\omega_{n0}}],\notag\\
 &&&\delta \textbf{j}(\textbf{r})\equiv\delta\langle \textbf{j} (\textbf{r})\rangle\notag\\
 &&&=\xi e^{-i(\omega+i\eta)t}[\frac{\langle0|\textbf{j}(\textbf{r})|n\rangle\langle n|\psi^{\dag}_{\textbf{q}}|0\rangle}{\omega+i\eta-\omega_{n0}}-\frac{\langle0|\psi^{\dag}_{\textbf{q}}|n\rangle\langle n|\textbf{j}(\textbf{r})|0\rangle}{\omega+i\eta+\omega_{n0}}]\notag\\
 &&&+\xi^* e^{i(\omega-i\eta)t}[\frac{\langle0|\psi_{\textbf{q}}|n\rangle\langle n|\textbf{j}(\textbf{r})|0\rangle}{\omega-i\eta-\omega_{n0}}-\frac{\langle0|\textbf{j}(\textbf{r})|n\rangle\langle n|\psi_{\textbf{q}}|0\rangle}{\omega-i\eta+\omega_{n0}}].\notag\\
\end{align}
In the following, we assume the system has translational invariance. So every eigenstate $|n\rangle$ has definite momentum $\textbf{q}_n$, e.g., $\textbf{P}|n\rangle=\textbf{q}_n|n\rangle$ with momentum operator $\textbf{P}=\sum_\textbf{k} \textbf{k}\psi_{\textbf{k}}^{\dag}\psi_\textbf{k}$.
On other hand, from commutation relations
\begin{align}
 &&[\textbf{P},\psi_{\textbf{q}}^{\dag}]|n\rangle=\{\textbf{P}\psi_{\textbf{q}}^{\dag}-\psi_{\textbf{q}}^{\dag}\textbf{P}\}|n\rangle=\textbf{q}\psi_{\textbf{q}}^{\dag}|n\rangle,\notag\\
 &&[\textbf{P},\psi_{\textbf{q}}]|n\rangle=\{\textbf{P}\psi_{\textbf{q}}-\psi_{\textbf{q}}\textbf{P}\}|n\rangle=-\textbf{q}\psi_{\textbf{q}}|n\rangle,
\end{align}
we see $\psi_{\textbf{q}}^{\dag}|n\rangle$ and $\psi_{\textbf{q}}|n\rangle$ also have definite momenta which are  $\textbf{q}_n+\textbf{q}$ and $\textbf{q}_n-\textbf{q}$, respectively.
Using $\psi(\textbf{r})=\sum_\textbf{q} \psi_\textbf{q} e^{i\textbf{q}\cdot\textbf{r}}$ and $\langle 0|\psi_{\textbf{q}'}|n\rangle\langle n|\psi_{\textbf{q}}^{\dag}|0\rangle=\delta_{\textbf{q},\textbf{q}'}|\langle 0|\psi_\textbf{q}|n\rangle|^2$, $\langle 0|\psi_{\textbf{q}}^{\dag}|n\rangle\langle n|\psi_{\textbf{q}'}|0\rangle=\delta_{\textbf{q},\textbf{q}'}|\langle 0|\psi_{\textbf{q}}^{\dag}|n\rangle|^2$, $\langle 0|\psi_{\textbf{q}}|n\rangle\langle n|\psi_{\textbf{q}'}|0\rangle=\delta_{\textbf{q},-\textbf{q}'}\langle 0|\psi_{\textbf{q}}|n\rangle\langle n|\psi_{-\textbf{q}}|0\rangle$, $\langle 0|\psi_{\textbf{q}'}|n\rangle\langle n|\psi_{\textbf{q}}|0\rangle=\delta_{\textbf{q},-\textbf{q}'}\langle 0|\psi_{-\textbf{q}}|n\rangle\langle n|\psi_{\textbf{q}}|0\rangle$,
the variation of order parameter can be written as
\begin{align}
 &&&\delta\langle\psi(\textbf{r})\rangle\notag\\
 &&&=\xi e^{i\textbf{q}\cdot \textbf{r}-i(\omega+i\eta)t} G(\textbf{q},\omega+i\eta)+\xi^* e^{-i\textbf{q}\cdot \textbf{r}+i(\omega-i\eta)t}F(\textbf{q},\omega-i\eta),\notag\\
 &&&G(\textbf{q},\omega+i\eta)=\sum_n[\frac{|\langle0|\psi_\textbf{q}|n\rangle|^2}{\omega+i\eta-\omega_{n0}}
 -\frac{|\langle0|\psi^{\dag}_\textbf{q}|n\rangle|^2}{\omega+i\eta+\omega_{n0}}],\notag\\
 &&&F(\textbf{q},\omega-i\eta)=\sum_n[\frac{\langle0|\psi_{\textbf{q}}|n\rangle\langle n|\psi_{-\textbf{q}}|0\rangle}{\omega-i\eta-\omega_{n0}}-\frac{\langle0|\psi_{-\textbf{q}}|n\rangle\langle n|\psi_{\textbf{q}}|0\rangle}{\omega-i\eta+\omega_{n0}}].
\end{align}

Taking zero-frequency of $\omega\pm i\eta =0$,
\begin{align}
 &&\delta\langle\psi(\textbf{r})\rangle=\xi e^{i\textbf{q}\cdot \textbf{r}} G(\textbf{q},0)+\xi^* e^{-i\textbf{q}\cdot \textbf{r}}F(\textbf{q},0).
\end{align}
For single component interacting bosons,  the low energy excitation is gapless phonon \cite{Landau1941,Gavoret}. It can be shown $F(\textbf{q},0)=-G(\textbf{q},0)$ as $q\rightarrow0$ \cite{Bogoliubov,Lifshitz}, so finally
\begin{align}
&&\delta\langle\psi(\textbf{r})\rangle=G(\textbf{q},0)[\xi e^{i\textbf{q}\cdot \textbf{r}} -\xi^* e^{-i\textbf{q}\cdot \textbf{r}}]\notag\\
&&=2i\alpha G(\textbf{q},0)sin(\textbf{q}\cdot \textbf{r}+\phi),\notag\\
\end{align}
where we take $\xi\equiv\alpha e^{i\phi}$.

Similarly, there is relationship
\begin{align}
 &&[\textbf{P}_i,\textbf{j}_{\textbf{q}j}]=-\textbf{q}_i \textbf{j}_{\textbf{q}j},
\end{align}
where $i,j=x,y,z$ and current fluctuation operator $\textbf{j}_{\textbf{q}}=\sum_\textbf{k} [\textbf{k}+\textbf{q}/2]\psi^{\dag}_{\textbf{k}}\psi_{\textbf{k}+\textbf{q}}$. $\textbf{j}_{\textbf{q}}|n\rangle$ also has definite momentum $\textbf{q}_n-\textbf{q}$. Using the fact of $\textbf{j}(\textbf{r})=\sum_\textbf{k} \textbf{j}_\textbf{k} e^{i\textbf{k}\cdot\textbf{r}}$ and $\langle 0|\textbf{j}_{\textbf{q}'}|n\rangle\langle n|\psi_{\textbf{q}}^{\dag}|0\rangle=\delta_{\textbf{q},\textbf{q}'}\langle 0|\textbf{j}_\textbf{q}|n\rangle\langle n|\psi_{\textbf{q}}^{\dag}|0\rangle$, $\langle 0|\psi_{\textbf{q}}^{\dag}|n\rangle\langle n|\textbf{j}_{\textbf{q}'}|0\rangle=\delta_{\textbf{q},\textbf{q}'}\langle 0|\psi_{\textbf{q}}^{\dag}|n\rangle\langle n|\textbf{j}_{\textbf{q}}|0\rangle$, the variation of current is
\begin{align}
\delta \textbf{j}(\textbf{r})=\xi e^{i\textbf{q}\cdot \textbf{r}-i(\omega+i\eta)t}\textbf{B}(\textbf{q},\omega+i\eta)+hc,
\end{align}
where $hc$ denotes Hermitian (complex) conjugate and
 \begin{align}
\textbf{B}(\textbf{q},\omega+i\eta)\equiv\sum_n[\frac{\langle0|\textbf{j}_q|n\rangle\langle n|\psi^{\dag}_{\textbf{q}}|0\rangle}{\omega+i\eta-\omega_{n0}}-\frac{\langle0|\psi^{\dag}_\textbf{q}|n\rangle\langle n|\textbf{j}_\textbf{q}|0\rangle}{\omega+i\eta+\omega_{n0}}].
\end{align}

When $\omega\pm i\eta=0$,
\begin{align}
\delta\textbf{ j}(\textbf{r})=[\xi e^{i\textbf{q}\cdot \textbf{r}}\textbf{B}(\textbf{q},0) +hc].
\end{align}


Using continuity equation, $\omega_{n0}(\rho_\textbf{q})_{0n}=\textbf{q}\cdot(\textbf{j}_{\textbf{q}})_{0n}$ and $\omega_{n0}(\rho_\textbf{q})_{n0}=-\textbf{q}\cdot(\textbf{j}_{\textbf{q}})_{n0}$, we can obtain
\begin{align}
&&\textbf{q}\cdot\textbf{B}(\textbf{q},0)=-\sum_n[\langle0|\rho_\textbf{q}|n\rangle\langle n|\psi^{\dag}_{\textbf{q}}|0\rangle-\langle0|\psi^{\dag}_\textbf{q}|n\rangle\langle n|\rho_\textbf{q}|0\rangle]\notag\\
&&=-[\langle0|\rho_\textbf{q},\psi^{\dag}_\textbf{q}]|0\rangle=-\langle0|\psi^{\dag}_{\textbf{q}=\textbf{0}}|0\rangle=-\langle\psi^\dag\rangle,
\end{align}
where density fluctuation operator $\rho_\textbf{q}=\sum_\textbf{k} \psi^{\dag}_{\textbf{k}}\psi_{\textbf{k}+\textbf{q}}$. So
\begin{align}
&&\textbf{q}\cdot \delta \textbf{j}(r)=-[\xi e^{i\textbf{q}\cdot \textbf{r}}\langle\psi^\dag\rangle+hc].
\label{yiban}
\end{align}
For isotropic system, further assuming $\textbf{q}\parallel \textbf{B}\propto \delta \textbf{j}$, so
\begin{align}
&\delta \textbf{j}(\textbf{r})=-\frac{\sqrt{n_0}\textbf{q}}{q^2}[\xi e^{i\textbf{q}\cdot \textbf{r}}+\xi^* e^{-i\textbf{q}\cdot \textbf{r}}]\notag\\
&=-\sqrt{n_0} \frac{\textbf{q}}{q^2} 2\alpha cos(\textbf{q}\cdot \textbf{r}+\phi)=-\frac{\sqrt{n_0}}{q^2} \frac{\vec{\nabla }\delta \langle \psi(\textbf{r})\rangle}{iG(\textbf{q},0)}.
\end{align}
Here we further use $\delta\langle\psi(\textbf{r})\rangle=i\langle\psi\rangle \delta \theta$, and then get
\begin{align}
&&\delta \textbf{j}(\textbf{r})= -\frac{\langle\psi^\dag\rangle}{q^2} \frac{\langle\psi\rangle\vec{\nabla }\delta \theta(\textbf{r})}{G(\textbf{q},0)}=-\frac{n_0}{q^2} \frac{ \delta \textbf{v}_s}{G(\textbf{q},0)},
\end{align}
where condensate density $n_0=\langle\psi^\dag\rangle\langle\psi\rangle$.
Using $\delta \textbf{j}(\textbf{r})\equiv\rho_s  \delta \textbf{v}_s$, the Josephson relation for usual single-component bosons is
\begin{align}
&&\rho_s=-lim_{q\rightarrow0}\frac{n_0}{q^2 G(\textbf{q},0)} .\label{Josephson}
\end{align}



\section{General Josephson relation for multiple-component Bosons }
For multiple-component Bosons (component number is arbitrary $n$), we need general perturbation Hamiltonian
\begin{align}
 H'&&&=\int d^3r\{e^{i(\textbf{q}\cdot \textbf{r}-\omega t)} \psi^{\dag}(r).\xi+e^{-i(\textbf{q}\cdot\textbf{ r}-\omega t)}\xi^{\dag} .\psi(r)\},\notag\\
 &&&=\psi^{\dag}_{\textbf{q}}.\xi e^{-i\omega t+\eta t}+\xi^{\dag}.\psi_{\textbf{q}}e^{i\omega t+\eta t},
\end{align}
where we introduce column vector $\psi(\textbf{r})=\{\psi_1(\textbf{r}),\psi_2(\textbf{r}),...,\psi_n(\textbf{r}) \}^\textbf{t}$ and $\xi=\{ \xi_1,\xi_2,...,\xi_n \}^\textbf{t}$.
Similarly, using the perturbation theory, we can get the variation of order parameter
 $\langle\psi(r)\rangle$
\begin{align}
 &&&\delta\langle\psi_{\sigma}(\textbf{r})\rangle=\sum_{\sigma'}[e^{i\textbf{q}\cdot \textbf{r}-i(\omega+i\eta)t}  G_{\sigma,\sigma'}(\textbf{q},\omega+i\eta)\xi_{\sigma'}\notag\\
 &&&+e^{-i\textbf{q}\cdot \textbf{r}+i(\omega-i\eta)t} F_{\sigma',\sigma}(\textbf{q},\omega-i\eta) \xi_{\sigma'}^{*}],\notag\\
 &&&G_{\sigma,\sigma'}(\textbf{q},\omega+i\eta)\notag\\
 &&&=\sum_n[\frac{\langle0|\psi_{\sigma \textbf{q}}|n\rangle\langle n|\psi^{\dag}_{\sigma' \textbf{q}}|0\rangle}{\omega+i\eta-\omega_{n0}}-\frac{\langle0|\psi^{\dag}_{\sigma' \textbf{q}}|n\rangle\langle n|\psi_{\sigma \textbf{q}}|0\rangle}{\omega+i\eta+\omega_{n0}}],\notag\\
 &&&F_{\sigma',\sigma}(\textbf{q},\omega-i\eta)\notag\\
 &&&=\sum_n[\frac{\langle0|\psi_{\sigma',\textbf{q}}|n\rangle\langle n|\psi_{\sigma,-\textbf{q}}|0\rangle}{\omega-i\eta-\omega_{n0}}-\frac{\langle0|\psi_{\sigma,-\textbf{q}}|n\rangle\langle n|\psi_{\sigma',\textbf{q}}|0\rangle}{\omega-i\eta+\omega_{n0}}].
 \label{definition}
\end{align}

For multi-component bosons, we assume the system Hamiltonian has symmetry which is described by some continuous group.
The group has several \emph{Hermitian} infinitesimal generators $\{T^\alpha\}$ with $\alpha=0,1,2,...,m-1$, where particularly, $T^0=I_{n\times n}$ (unit matrix) denotes the generator for  phase $U(1)$ group in bosons. In the derivation of Josephson relation, we use the continuity equation for particle number in above section.
In fact, we can investigate
 the continuity equations for general conserved charges, i.e.,
\begin{align}
&\omega_{n0}(\rho^{\alpha}_{\textbf{q}})_{0n}=\textbf{q}\cdot(\textbf{j}^{\alpha}_{\textbf{q}})_{0n},\notag\\
 &\omega_{n0}(\rho^{\alpha}_{\textbf{q}})_{n0}=-\textbf{q}\cdot(\textbf{j}^{\alpha}_{\textbf{q}})_{n0},
\end{align}
where
\begin{align}
&\rho^{\alpha}_\textbf{q}=\sum_{\textbf{k}\mu\nu}\psi^{\dag}_{\mu\textbf{k}}T^{\alpha}_{\mu\nu}\psi_{\nu\textbf{k}+\textbf{q}},\notag\\
&\textbf{j}^{\alpha}_{\textbf{q}}=\sum_{\textbf{k}\mu\nu}[\textbf{k}+\textbf{q}/2]\psi^{\dag}_{\mu\textbf{k}}T^{\alpha}_{\mu\nu}\psi_{\nu\textbf{k}+\textbf{q}}.
\end{align}

Similarly as above, one gets
\begin{align}
\delta \textbf{j}^\alpha(\textbf{r})=e^{i\textbf{q}\cdot\textbf{r}-i(\omega+i\eta)t} \textbf{B}^{\alpha}(q,\omega+i\eta).\xi+hc,
\end{align}
and
\begin{align}
 &&&\textbf{B}^{\alpha}_{\sigma'}(q,\omega+i\eta)\notag\\
 &&&=\sum_n[\frac{\langle0|\textbf{j}^{\alpha}_\textbf{q}|n\rangle\langle n|\psi_{\sigma'\textbf{q}}^{\dag}|0\rangle}{\omega+i\eta-\omega_{n0}}-\frac{\langle0|\psi_{\sigma'\textbf{q}}^{\dag}|n\rangle\langle n|\textbf{j}^{\alpha}_\textbf{q}|0\rangle}{\omega+i\eta+\omega_{n0}}].\notag\\
\end{align}

When $\omega\pm i\eta= 0$,
\begin{align}
 &&&\delta\langle\psi(\textbf{r})\rangle=e^{i\textbf{q}\cdot\textbf{r}} G(\textbf{q},0).\xi+e^{-i\textbf{q}\cdot\textbf{r}}  F^\textbf{t}(\textbf{q},0).\xi^{*},\notag\\
 &&&\delta \textbf{j}^{\alpha}(\textbf{r})=e^{i\textbf{q}\cdot\textbf{r}} \textbf{B}^{\alpha}(\textbf{q},0).\xi+hc.
 \label{current}
\end{align}
Similarly, using continuity equations and assuming $\textbf{q}\parallel \textbf{B}^{\alpha}\propto \delta \textbf{j}^{\alpha}$ for isotropic system, we get
\begin{align}
 \textbf{B}^{\alpha}(\textbf{q},0)=-\frac{\textbf{q}}{q^2}\{\langle\psi^{\dag}_{1}\rangle,\langle\psi^{\dag}_{2}\rangle,...,\langle\psi^{\dag}_{n}\rangle\}.T^{\alpha}.
\end{align}

Let's introduce $x(\textbf{r})=e^{i\textbf{q}\cdot\textbf{r}}\{\xi_{1},\xi_{2},...,\xi_{n} \}^\textbf{t}$, $\delta \langle\psi(\textbf{r})\rangle=\{\delta\langle\psi_{1}(\textbf{r})\rangle, \delta\langle\psi_{2}(\textbf{r})\rangle,...,\delta\langle\psi_{n}(\textbf{r})\rangle\}^\textbf{t}$,  we can rewrite the above equations as
\begin{align}
&&\left( \!\!\!                
 \begin{array}{cccc}   
   \delta \langle\psi(\textbf{r})\rangle    \\  
   \delta \langle\psi(\textbf{r})\rangle^{*} \\  
\end{array}\!\!\!\right)\!\!=\!\!\left( \!\!\!                
 \begin{array}{cccc}   
   G(\textbf{q},0)  & F^\textbf{t}(\textbf{q},0) \\  
   F^{t*}(\textbf{q},0)& G^{*}(\textbf{q},0)\\  
\end{array}\!\!\!\right).\left( \!\!\!                
 \begin{array}{cccc}   
   x   \\  
   x^* \\  
\end{array}\!\!\!\right)\equiv\textbf{G}.\left( \!\!\!                
 \begin{array}{cccc}   
   x   \\  
   x^* \\  
\end{array}\!\!\!\right),
\label{basic}
\end{align}

and
\begin{align}
&&\delta \textbf{j}^{\alpha}(\textbf{r})\!\!=\!\!-\frac{\textbf{q}}{q^2}\left(                
 \begin{array}{cccc}   
   \langle\psi\rangle^{t*}  &\langle\psi\rangle^{t} \\  
\end{array}\right).\!\!\left( \!\!\!                
 \begin{array}{cccc}   
   T^\alpha  & 0 \\  
   0& T^{\alpha*}\\  
\end{array}\!\!\!\right).\left(                 
 \begin{array}{cccc}   
   x   \\  
   x^* \\  
\end{array}\!\!\!\right),
\label{basic2}
\end{align}
where we define coefficient matrix $\textbf{G}\equiv\!\!\left( \!\!\!                
 \begin{array}{cccc}   
   G(\textbf{q},0)  & F^\textbf{t}(\textbf{q},0) \\  
   F^{t*}(\textbf{q},0)& G^{*}(\textbf{q},0)\\  
\end{array}\!\!\!\right)$ and one should not confuse with normal Green's function $G(\textbf{q},0)$. The two equations (\ref{basic}), (\ref{basic2}) are basic equations in this work. In the following, most discussions are based on them.

In the following, we need to generalize the eq. (\ref{deltaphi}) to describe general spin currents in multiple-component case. For such purpose,  we assume all the order parameters undergo a specific phase variation generated by $T^\beta$, i.e.,
\begin{align}\label{variation}
&\left(               
 \begin{array}{cccc}   
   \delta \langle\psi(\textbf{r})\rangle    \\  
   \delta \langle\psi(\textbf{r})\rangle^{*} \\  
\end{array}\right)\!\!=\!\!\left( \!\!             
 \begin{array}{cccc}   
   e^{i\delta\theta^{\beta}T^\beta}  & 0 \\  
  0&  e^{-i\delta\theta^{\beta}T^{\beta*}}\\  
\end{array}\!\!\right).\left(  \!\!              
 \begin{array}{cccc}   
   \langle\psi\rangle   \\  
   \langle\psi\rangle^* \\  
\end{array}\!\!\right)\!\!-\!\!\left( \!\!               
 \begin{array}{cccc}   
   \langle\psi\rangle   \\  
   \langle\psi\rangle^* \\  
\end{array}\!\!\right)\notag\\
&=i\delta\theta^{\beta}\left(              
 \begin{array}{cccc}   
   T^\beta  & 0 \\  
  0& -T^{\beta*}\\  
\end{array}\!\!\right).\left( \!\!               
 \begin{array}{cccc}   
   \langle\psi\rangle   \\  
   \langle\psi\rangle^* \\  
\end{array}\!\!\right).
\end{align}

If $\textbf{G}$ has inverse ($Det|\textbf{G}|\neq0$),
 using $\textbf{q} x(\textbf{r})=-i \vec{\nabla } x$, $\textbf{q} x^*(\textbf{r})=i \vec{\nabla } x^*$ and eqs. (\ref{basic},\ref{basic2})
we get
\begin{eqnarray}
&& \delta \textbf{j}^{\alpha}(\textbf{r})=-\frac{\vec{\nabla}\delta \theta^{\beta}(\textbf{r})}{q^2}[\langle\psi^{t*}\rangle,\langle\psi\rangle^{t}]\notag\\
&&.\!\!\left( \!\!\!                
 \begin{array}{cccc}   
  T^{\alpha}  & 0 \\  
   0&  -T^{\alpha*}\\  
\end{array}\!\!\!\right).\textbf{G}^{-1}.\!\!\left( \!\!\!                
 \begin{array}{cccc}   
  T^{\beta}  & 0 \\  
   0&  -T^{\beta*}\\  
\end{array}\!\!\!\right).\left(
 \begin{array}{cccc}   
   \langle\psi\rangle \\  
   \langle\psi\rangle^*\\  
\end{array}\right),\notag\\
&&=-\frac{\delta \textbf{v}^{\beta}_s}{q^2}[\langle\psi\rangle^{t*},\langle\psi\rangle^t].\notag\\
&&.\!\!\left( \!\!\!                
 \begin{array}{cccc}   
  T^{\alpha}  & 0 \\  
   0&  -T^{\alpha*}\\  
\end{array}\!\!\!\right).\textbf{G}^{-1}.\!\!\left( \!\!\!                
 \begin{array}{cccc}   
  T^{\beta}  & 0 \\  
   0&  -T^{\beta*}\\  
\end{array}\!\!\!\right).\left(
 \begin{array}{cccc}   
   \langle\psi\rangle \\  
   \langle\psi\rangle^*\\  
\end{array}\right),
\end{eqnarray}
where  $\delta \textbf{v}^{\beta}_s\equiv \nabla \delta \theta^{\beta}$ is superfluid velocity for $T^\beta$.
  Using  eq. (\ref{deltaj}), i.e., $\delta \textbf{j}^{\alpha}(\textbf{r})\equiv\rho^{\alpha\beta}_s \delta \textbf{v}^{\beta}_s$, we get generalized Josephson relation for superfluid density
\begin{eqnarray}
&& \rho^{\alpha\beta}_s\!\!=\!\!lim_{q\rightarrow0}-\frac{1}{q^2}[\langle\psi\rangle^{t*},\langle\psi\rangle^t].\!\!\left( \!\!\!                
 \begin{array}{cccc}   
  T^{\alpha}  & 0 \\  
   0&  -T^{\alpha*}\\  
\end{array}\!\!\!\right)\notag\\
&&.\textbf{G}^{-1}.\!\!\left( \!\!\!                
 \begin{array}{cccc}   
  T^{\beta}  & 0 \\  
   0&  -T^{\beta*}\\  
\end{array}\!\!\!\right).\left(
 \begin{array}{cccc}   
   \langle\psi\rangle \\  
   \langle\psi\rangle^*\\  
\end{array}\right).
\end{eqnarray}

In isotropic case, the superfluid density in orbital space is like a scalar, i.e., $\rho_s=\textbf{diag}\{\rho^{\alpha\beta}_s ,\rho^{\alpha\beta}_s ,\rho^{\alpha\beta}_s \}$ which does not depend on $\textbf{q}$ 's direction.
For generally anisotropic system (for example, the spin-orbital coupled BEC \cite{normaldensity}), the induced current $\delta \textbf{j}^{\alpha}$ can be expressed in terms of vector $\textbf{q}$ and a second order tensor $m$ (generally $\delta\textbf{ j}^{\alpha}$ is not parallel to $\textbf{q}$ any more), namely
\begin{align}\label{tensor}
\delta \textbf{j}^{\alpha}_{i}(\textbf{r})=\sum_j m_{ij}\textbf{q}_j.
\end{align}
 In order to get Josephson relation in anisotropic system, we define the superfluid density $\rho^{\alpha\beta}_{s}(\hat{q})$ and superfluid velocity $\hat{q}\cdot \delta \textbf{v}^{\beta}_s$ along $\hat{q}$ direction, i.e., $\delta \textbf{j}^{\alpha}_{\hat{q}}(\textbf{r})\equiv\hat{q}\cdot \delta \textbf{j}^{\alpha}(\textbf{r})\equiv\rho^{\alpha\beta}_{s}(\hat{q}) (\hat{q}\cdot \delta \textbf{v}^{\beta}_s)$, $\hat{q}\equiv \textbf{q}/q$ is unit vector along $\textbf{q}$ direction.  From eq. (\ref{current},\ref{tensor}), we get
\begin{align}
&&&\delta \textbf{j}^{\alpha}_{\hat{q}}(\textbf{r})\equiv\hat{q}\cdot \delta \textbf{j}^{\alpha}(\textbf{r})=\sum_{ij}m_{ij}\hat{q}_i \hat{q}_j\notag\\
&&&=-\frac{q}{q^2}\left(                
 \begin{array}{cccc}   
   \langle\psi\rangle^{t*}  &\langle\psi\rangle^{t} \\  
\end{array}\right).\!\!\left( \!\!\!                
 \begin{array}{cccc}   
   T^\alpha  & 0 \\  
   0& T^{\alpha*}\\  
\end{array}\!\!\!\right).\left(                 
 \begin{array}{cccc}   
   x   \\  
   x^* \\  
\end{array}\!\!\!\right),
\end{align}
which is exactly similar to eq.(\ref{basic2}). Taking $q x(\textbf{r})=-i \hat{q}\cdot\vec{\nabla } x$, $q x^*(\textbf{r})=i \hat{q}\cdot\vec{\nabla } x^*$ and eq.(\ref{basic}) into account, so the following discussions for anisotropic case are same with that in the isotropic case.
Then, from the superfluid density along arbitrary direction $\hat{q}$, i.e.,
\begin{align}
\rho^{\alpha\beta}_s(\hat{q})=\sum_{ij}\rho^{\alpha\beta}_{s;ij} \hat{q}_i \hat{q}_j,
\end{align}
it is not difficult to construct second order tensor $\rho^{\alpha\beta}_{s;ij}$.

The above discusses show that the superfluid density is generally second order tensor in internal spin and orbital space, respectively.
When several superfluid velocities coexist, the super-current can be written as  \cite{Leggett}
\begin{align}
\textbf{j}^{\alpha}_{i}=\sum_{\beta j}\rho^{\alpha\beta}_{s;ij}\textbf{v}^{\beta}_{s;j},
\end{align}
where index $\{\alpha\beta\}$ is for internal spin space and $\{ij\}$ is for external orbit space.
We should remark that because the above results are obtained from the perturbation theory,  the states for above superfluid currents should be not far from the ground states or thermodynamic equilibrium states.

\section{Discussions }
In the above derivation, in order to get the Josephson relation, we use a crucial assumption of $\delta \langle\psi_{\sigma}(\textbf{r})\rangle=i \delta \theta^{\beta}(\textbf{r})\sum_{\sigma'}T^{\beta}_{\sigma\sigma'}\langle \psi_{\sigma'}\rangle$ for variations of order parameters.  Such an assumption must be consistent with the linear equation
\begin{align}
\left( \!\!\!                
 \begin{array}{cccc}   
   \delta \langle\psi(\textbf{r})\rangle    \\  
   \delta \langle\psi(\textbf{r})\rangle^*\\  
\end{array}\!\!\!\right)\!\!=\!\!\left( \!\!\!                
 \begin{array}{cccc}   
   G(\textbf{q},0)  &F^t(\textbf{q},0) \\  
   F^{t*}(\textbf{q},0) & G^{*}(\textbf{q},0)\\  
\end{array}\!\!\!\right).\left( \!\!\!                
 \begin{array}{cccc}   
   x   \\  
   x^* \\  
\end{array}\!\!\!\right).
\end{align}
In other word, the above equation should have at least one solution  of $x$ in terms of $\delta \theta^{\beta}$. According to the number of solutions, we have several different cases.

\subsection{ unique solution when $Det |\textbf{G}|\neq0$ }
When $Det |\textbf{G}|\neq0$, and coefficient matrix $\textbf{G}$ has inverse, we always have unique solution for arbitrary order prameter variation $\delta\langle\psi\rangle$.
An example is the non-interacting Bose gas where the anomalous Green's function $F\equiv0$ \cite{Lifshitz}. The normal Green's  function $G(\textbf{q},\omega)=1/(\omega-q^2/2+\mu)$, here $\mu$ is chemical potential. So the superfluid density for non-interacting Bose-Einstein condensate ($\mu=0$ for condensate) is condensate density $n_0$
\begin{eqnarray}
& \rho^{00}_s=\frac{-1}{q^2}(\langle\psi\rangle^{t*},\langle\psi\rangle^{t}).\sigma_z.\textbf{G}^{-1}.\sigma_z.\left(
 \begin{array}{cccc}   
   \langle\psi\rangle \\  
   \langle\psi\rangle^*\\  
\end{array}\right)\notag\\
&=\frac{-1}{q^2}(\langle\psi\rangle^{t*},\langle\psi\rangle^{t}).\left( \!\!\!                
 \begin{array}{cccc}   
  - q^2/2  &0 \\  
   0 & - q^2/2\\  
\end{array}\!\!\!\right).\left(
 \begin{array}{cccc}   
   \langle\psi\rangle \\  
   \langle\psi\rangle^*\\  
\end{array}\right)\notag\\
&=n_0,
\end{eqnarray}
where order parameter $\langle\psi\rangle=\sqrt{n_0}$ and the generator $T^0=1$ for phase $U(1)$ group in single component case.
This formula resolves an evident contradiction that if we apply usual Josephson's relation of Eq. (\ref{Josephson}) to non-interacting Bose gas, we only get an half of particles which has superfluidity.

For two-component bosons, we assume the order parameter is $\langle\psi\rangle=\{\sqrt{n_0}/\sqrt{2},\sqrt{n_0}/\sqrt{2}\}^t$.
 The system has $U(2)$ symmetry. The group generators are $T^{0}=I_{2\times2}$, $T^{1}=\sigma_x$, $T^{2}=\sigma_y$ and $T^{3}=\sigma_z$ .
 The non-vanishing superfluid density matrix elements are
\begin{eqnarray}
&\rho^{\alpha\alpha}_s=n_0,\notag\\
&\rho^{01}_s=\rho^{10}_s=n_0.
\end{eqnarray}

\subsection{ infinite  solutions when $Det |\textbf{G}|=0$  }

According to the theory of linear equations, when $Det |\textbf{G}|=0$, and  the rank of augmented matrix of $\textbf{G}: b$ ($b=\left( \!\!\!
 \begin{array}{cccc}   
   iT^{\beta}.\langle\psi\rangle    \\  
    -iT^{\beta*}.\langle\psi\rangle^* \\  
\end{array}\!\!\!\right)$)  is equal to rank of coefficient matrix $\textbf{G}$, we have infinite solutions.
In the following, we list several examples.

\subsubsection{Rank(\textbf{G})=1 }
When system has unique gapless excitation near ground state, e.g., phonon, the Josephson relation can be extended to multi-component system by phase operator method. In such case, generally speaking, the rank of coefficient matrix  $Rank(\textbf{G})=1$.
Here we know near the ground states, the phonon's excitation corresponds to total density oscillation. Due to presence of condensate, the density oscillation would couple with condensate phase oscillation.  Furthermore, all the components should share a common phase variation, i.e., $\delta \theta_\sigma(\textbf{r})=\delta\theta(\textbf{r})$. On the other hand, near ground state, the field operators can be expressed in terms of  phase operator \cite{Lifshitz}
\begin{align}
\psi_{\sigma}(\textbf{r})=\sqrt{n_{0,\sigma}}e^{i\theta_{\sigma}(\textbf{r})},
\end{align}
where $n_{0,\sigma}$ is condensate density of $\sigma-th$ component and condensate density is given by $n_0=\sum_{\sigma=1}^{n} n_{0,\sigma}$.
For long wave length phonon, the variation of phase is small (the variations of amplitudes can be neglected), so we can get
\begin{align}
\delta\psi_{\sigma}(\textbf{r})=\sqrt{n_{0,\sigma}}\delta e^{i\theta_{\sigma}(\textbf{r})}=i\langle\psi_\sigma\rangle\delta \theta(\textbf{r}).
\end{align}
From the above equation, we get the field operators in momentum space
\begin{align}
&\psi_{\sigma,\textbf{q}}=i\langle\psi_\sigma\rangle\theta_{\textbf{q}},\notag\\
&\psi^{\dag}_{\sigma,\textbf{q}}=-i\langle\psi^{\dag}_{\sigma}\rangle\theta^{\dag}_{\textbf{q}}=-i\langle\psi^{\dag}_{\sigma}\rangle\theta_{-\textbf{q}}.\notag\\
&\psi_{\sigma,-\textbf{q}}=i\langle\psi_\sigma\rangle\theta_{-\textbf{q}},\notag\\
&\psi^{\dag}_{\sigma,-\textbf{q}}=-i\langle\psi^{\dag}_{\sigma}\rangle\theta^{\dag}_{-\textbf{q}}=-i\langle\psi^{\dag}_{\sigma}\rangle\theta_{\textbf{q}},
\end{align}
where we use  $\theta^{\dag}_{\textbf{q}}=\theta_{-\textbf{q}}$ for real phase field $\theta(\textbf{r})$ .
From definitions of the $G$ and $F$ in eq.(\ref{definition}), we get
\begin{align}
 &G_{\sigma,\sigma'}(\textbf{q},0)=-Z\langle\psi_\sigma\rangle\langle\psi^{\dag}_{\sigma'}\rangle,\notag\\ &F_{\sigma',\sigma}(\textbf{q},0)=Z\langle\psi_{\sigma'}\rangle\langle\psi_{\sigma}\rangle ,
\end{align}
where $Z\equiv\sum_n[\frac{\langle0|\theta_{\textbf{q}}|n\rangle\langle n|\theta_{-\textbf{q}}|0\rangle}{\omega_{n0}}+\frac{\langle0|\theta_{-\textbf{q}}|n\rangle\langle n|\theta_{\textbf{q}}|0\rangle}{\omega_{n0}}]\geq0$ is a  real number. Consequently, the variations for order parameters can be assumed as $\delta\langle\psi_{\sigma}\rangle=i\langle\psi_{\sigma}\rangle\delta\theta$. For such variations, one can verify $Det|\textbf{G}|=0$ and the rank(\textbf{G})=rank(\textbf{G}:b)$=1<2n$, so we have infinite solutions in eq. (\ref{basic}).

From eqs.(\ref{basic}) and (\ref{basic2}), we get
\begin{align}
 &&&i\delta\theta(\textbf{r})=-Z\sum_\sigma [\langle \psi^{\dag}_{\sigma}\rangle x_\sigma-\langle\psi_\sigma\rangle x^{*}_{\sigma}]\notag\\
 &&&=-Z\sum_\sigma 2i\alpha_\sigma sin(\textbf{q}\cdot\textbf{r}+\phi_\sigma),\notag\\
 &&& \delta \textbf{j}(\textbf{r})=-\frac{\textbf{q}}{q^2}\sum_{\sigma}[\langle \psi^{\dag}_{\sigma}\rangle x_\sigma+\langle\psi_\sigma\rangle x^{*}_{\sigma}],\notag\\
 &&&=-\frac{\textbf{q}}{q^2}\sum_{\sigma}[2\alpha_\sigma cos(\textbf{q}\cdot\textbf{r}+\phi_\sigma)]=\frac{\vec{\nabla} \delta\theta(\textbf{r})}{q^2Z}=\frac{\delta \textbf{v}_s}{q^2Z},
\end{align}
where we take $\langle \psi^{\dag}_{\sigma}\rangle \xi_\sigma\equiv\alpha_\sigma e^{i\phi_\sigma}$ and use $(\langle\psi_\sigma\rangle)^*=\langle\psi^{\dag}_\sigma\rangle$.
So the superfluid density is
\begin{align}
\rho^{00}_s=\frac{1}{q^2Z}.
\end{align}

On the other hand, we know $\sum_\sigma \langle\psi_{\sigma}^{\dag}\rangle\langle\psi_\sigma\rangle=n_0$,
\begin{align}
tr G(\textbf{q},0)\equiv\sum_\sigma G_{\sigma,\sigma}(\textbf{q},0)=-Z n_0.
\end{align}
 So finally we get the Josephson relation for $Rank(\textbf{G})=1$
\begin{align}
\rho^{00}_s=-lim_{q\rightarrow0}\frac{n_0}{q^2 trG(\textbf{q},0)},
\end{align}
where $G(\textbf{q},0)$ is normal Green's function at zero-frequency. When the number of component $n=1$, the above equation is reduced to the usual Josephson relation for single component bosons (see eq. (\ref{Josephson})). In the above discuss,  we assume the symmetry group for Hamiltonian is phase $U(1)$ group whose unique generator is $T^0=I_{n\times n}$.
The above formula for superfluid density can be applied to two-component spin-orbit coupled BEC \cite{Lin1} . Within Bogoliubov approximation, the above results for superfluid density are consistent with the results from current-current correlation calculations \cite{normaldensity,zhangyicai}.

\subsubsection{Two-component interacting bosons}
Another example for $Det|\textbf{G}|=0$  is the two component bosons.
The Hamiltonian for two component bosons \cite{Pitaevskii,Pethick} is
\begin{eqnarray}\label{hamiltonian0}
     &&H=\int d^3\textbf{r}\sum_{\sigma=1,2}\psi_{\alpha}^{\dag}[\frac{p^{2}}{2m}-\mu]\psi_{\alpha}+V_{int} ,\notag\\
     &&V_{int}=\frac{1}{2}\int d^3\textbf{r}_1d^3\textbf{r}_2\notag\\
     &&\times\!\!\sum_{\alpha,\beta=1,2}\!\!\psi_{\alpha}^\dag(\textbf{r}_1)\psi_{\beta}^\dag(\textbf{r}_2)V_{\alpha\beta }(|\textbf{r}_1-\textbf{r}_2|)\psi_{\beta}(\textbf{r}_2)\psi_{\alpha}(\textbf{r}_1),\notag\\
\end{eqnarray}
where $V_{\alpha\beta}$ are interaction potentials between atoms, and we assume $V_{\alpha\beta}=V_{\beta\alpha}$ and are real. $\psi_{1(2)}$ are two component bosonic field operators. $m$ is atomic mass and $\mu$ is chemical potential. In the following, we set mass $m=\hbar=1$.

Here we generalize perturbation theory of quantum field \cite{Abrikosov, Fetter} to two-component bosons (see Appendix \textbf{A}) .
When Bose-Einstein condensation occurs in bosonic system, the field operators would take non-vanishing mean values, e.g.,
\begin{align}\label{eqn1}
  \langle \psi\rangle = \left(
 \begin{array}{cccc}   
   \langle \psi_1 \rangle \\  
   \langle \psi_2 \rangle\\  
\end{array}\right)= \left(
 \begin{array}{cccc}   
   \xi_{1} \\  
   \xi_{2}\\  
\end{array}\right)=\left(
 \begin{array}{cccc}   
   \sqrt{n_{1}} \\  
   \sqrt{n_{2}}\\  
\end{array}\right),
\end{align}
where $n_{\sigma=1,2}$ is condensate density for spin-up and spin-down component and $n_{1}+n_{2}=n_0$ gives the total condensate density.

 First of all,  similar to single component bosons, a generalized Hugenholtz-Pines relation \cite{Pines1959} hold (see Appendix \textbf{B}), i.e.,
\begin{align}
&&\!\!\left( \!\!\!                
 \begin{array}{cccc}   
   \mu  & 0 \\  
   0& \mu\\  
\end{array}\!\!\!\right)=\!\!\left( \!\!\!                
 \begin{array}{cccc}   
   \Sigma_{11,11}(\textbf{0},0)  & \Sigma_{11,12}(\textbf{0},0) \\  
   \Sigma_{11,21}(\textbf{0},0)& \Sigma_{11,22}(\textbf{0},0)\\  
\end{array}\!\!\!\right)\notag\\
&&-\!\!\left( \!\!\!                
 \begin{array}{cccc}   
   \Sigma_{20,11}(\textbf{0},0) & \Sigma_{20,12}(\textbf{0},0) \\  
   \Sigma_{20,21}(\textbf{0},0)& \Sigma_{20,22}(\textbf{0},0)\\  
\end{array}\!\!\!\right),
\end{align}
where $\Sigma_{11}(\textbf{0},0)$ ($\Sigma_{20}(\textbf{0},0)$) are $2\times2$ normal (anomalous) self-energy matrix at zero momentum-frequency, the second pair of subscripts are spin-indexes  for external non-condensate particle lines.

 In the Appendix \textbf{C}, for general interaction strengthes in Hamiltonian eq. (\ref{hamiltonian0}), we show there are two phonon's excitations with linear dispersion
  \begin{align}
&\omega_1(q)=s_1q, \notag\\
&\omega_2(q)=s_2q,
\end{align}
as $q\rightarrow0$.
At the zero-frequency of $\omega\pm i\eta=0$ and $q\rightarrow0$, we find the leading terms of normal (anomalous) Green's function matrix elements  are real and satisfy
\begin{align}
F_{\alpha \beta}(\textbf{q},0)=-G_{\alpha \beta}(\textbf{q},0)\propto 1/q^2.
\end{align}
 The above result is very similar to single-component case where $F(\textbf{q},0)=-G(\textbf{q},0)$.

For general case of $V_{11}\neq V_{12}\neq V_{22}$, the system has phase $U(1)$ and spin rotation symmetries generated by $T^0$ and $T^3=\sigma_z$.
 For both variations of $\delta\langle\psi\rangle$  specified by $T^0$ and $T^3$ in Eq. (\ref{variation}),  one can verify that $Det|\textbf{G}|=0$ , the rank(\textbf{G})=rank(\textbf{G}:b)$=2<4$, we have infinite solutions in eq. (\ref{basic}).

 From these results and eqs. (\ref{basic}), (\ref{basic2}), we can get the Josephson relation for two component bosons immediately
 \begin{eqnarray}
&& \rho^{00}_s=lim_{q\rightarrow0}\frac{-1}{q^2}(\langle\psi_{1}\rangle^{*},\langle\psi_{2}\rangle^{*}).G^{-1}(\textbf{q},0).\left(\!\!
 \begin{array}{cccc}   
   \langle\psi_1\rangle \\  
    \langle\psi_2\rangle \\  
\end{array}\!\!\right),\notag\\
&& \rho^{03}_s\notag\\
&&=lim_{q\rightarrow0}\frac{-1}{q^2}(\langle\psi_{1}\rangle^{*},\langle\psi_{2}\rangle^{*}).G^{-1}(\textbf{q},0).\sigma_z.\left(\!\!
 \begin{array}{cccc}   
   \langle\psi_1\rangle \\  
    \langle\psi_2\rangle \\  
\end{array}\!\!\right),\notag\\
&& \rho^{30}_s\notag\\
&&=lim_{q\rightarrow0}\frac{-1}{q^2}(\langle\psi_{1}\rangle^{*},\langle\psi_{2}\rangle^{*}).\sigma_z.G^{-1}(\textbf{q},0).\left(\!\!
 \begin{array}{cccc}   
   \langle\psi_1\rangle \\  
    \langle\psi_2\rangle \\  
\end{array}\!\!\right),\notag\\
&&\rho^{33}_{s}\notag\\
&&=lim_{q\rightarrow0}\frac{-1}{q^2}(\langle\psi_{1}\rangle^{*},\langle\psi_{2}\rangle^{*}).\sigma_z.G^{-1}(\textbf{q},0).\sigma_z.\left(\!\!
 \begin{array}{cccc}   
   \langle\psi_1\rangle \\  
    \langle\psi_2\rangle \\  
\end{array}\!\!\right).\notag\\
\end{eqnarray}
Here $\rho^{03}_s=\rho^{03}_s$ is due to $G_{\alpha\beta}(\textbf{q},0)=G_{\beta\alpha}(\textbf{q},0)$ (see eq. (\ref{selfenergy2}) in Appendix \textbf{A}).

When interaction strengths are all equal ($V_{11}=V_{12}=V_{22}$), the system has $U(2)$ symmetry, the group generator are $T^0$, $T^1$, $T^2$ and $T^3$. Order parameter for bosons is also U(2) degeneracy.  Then we can take
\begin{eqnarray}\label{eqn1}
    \langle\psi\rangle=\frac{\sqrt{n_0}}{\sqrt{2}}\left(
 \begin{array}{cccc}   
   1 \\  
   1\\  
\end{array}\right),
\end{eqnarray}
for example without loss generality. Within Bogoliubov approximation, it is shown that  there exist two branches of  gapless excitations. One is linear phonon $\omega_1(q)\sim q$ as $q\rightarrow0$, the other one has quadratic dispersion $\omega_2(q)\sim q^2$ \cite{Goldstein,Pitaevskii}.

In the appendix \textbf{D}, we show existence of  above two types of excitations (one is linear phonon, another is quadratic dispersion) in $U(2)$ invariant interaction bosons for arbitrary orders of perturbation theory.
Furthermore, at the zero-frequency and as $q\rightarrow0$, the Green's functions satisfy
\begin{align}
&G_{11}(\textbf{q},0)=G_{22}(\textbf{q},0),\notag\\
&G_{12}(\textbf{q},0)=G_{21}(\textbf{q},0),\notag\\
&F_{11}(\textbf{q},0)=F_{22}(\textbf{q},0),\notag\\
&F_{12}(\textbf{q},0)=F_{21}(\textbf{q},0),\notag\\
&F_{11}(\textbf{q},0)+F_{12}(\textbf{q},0)\notag\\
&=-[G_{11}(\textbf{q},0)+G_{12}(\textbf{q},0)]\propto 1/q^2.
\label{guanxi}
\end{align}
One can verify $Det|\textbf{G}|=0$ and the rank(\textbf{G})=rank(\textbf{G}:b)$=3<4$, we also have infinite solutions in eq. (\ref{basic}).
Using the eqs.(\ref{basic}), (\ref{basic2}), we get non-vanishing superfluid densities for $U(2)$ invariant interaction bosons
 \begin{eqnarray}
&& \rho^{00}_s= \rho^{10}_s=\rho^{01}_s=\rho^{11}_s\notag\\
&&\!\!=\!\! lim_{q\rightarrow0}\frac{-1}{q^2}(\langle\psi_{1}\rangle^{*},\langle\psi_{2}\rangle^{*}).G^{-1}(\textbf{q},0).\!\!\left(\!\!
 \begin{array}{cccc}   
   \langle\psi_1\rangle \\  
    \langle\psi_2\rangle \\  
\end{array}\!\!\right)\!\!,\notag\\
&&= lim_{q\rightarrow0}\frac{-n_0}{q^2[G_{11}(\textbf{q},0)+G_{12}(\textbf{q},0)]},\notag\\
&& \rho^{22}_s= lim_{q\rightarrow0}\frac{-n_0}{q^2[G_{11}(\textbf{q},0)+F_{11}(\textbf{q},0)]},\notag\\
&& \rho^{33}_s= lim_{q\rightarrow0}\frac{n_0}{q^2[G_{12}(\textbf{q},0)+F_{11}(\textbf{q},0)]},\notag\\
&&= lim_{q\rightarrow0}\frac{-n_0}{q^2[G_{11}(\textbf{q},0)+F_{12}(\textbf{q},0)]}.
\end{eqnarray}
At the last step in above equation, we use eq.(\ref{guanxi}).

\section{summary}
In conclusions, we investigate the Josephson relation for general conserved charges in multiple-component bosons.
 When there are several conserved charges, generally speaking, the superfluid density is second order tensor in internal spin space.
From the linear response theory, we give a general formula to discuss the Josephson relation. Once the relations between the normal and anomalous Green's functions are known, based on the general formula, the Josephson relation can be established. To be specific,
when there is an unique gapless excitation, the Josephson relation is obtained by using phase operator method. For two component bosons, we show a generalized Hugenholtz-Pines relation hold and existence of two branches of phonon excitation. For $U(2)$ invariant interaction bosons, there exists one branch of quadratic dispersion excitations no matter how strong the interactions are.
Our work provides insights for understanding the superfluid properties of complex multiple-component bosonic system.

\acknowledgements
We thank Shizhong Zhang for useful discussions.
This work is supported by Hong Kong Research
Grants Council (General Research Fund, HKU 17318316 and Colaborative Research Fund, C6026-16W).
We also thank the supports of startup grant from Guangzhou University.
\appendix

 \section{General formalism for two-component bosons }
The Hamiltonian for two component Bosons is \cite{Pitaevskii,Pethick}
\begin{eqnarray}\label{hamiltonian}
     &&H=\int d^3\textbf{r}\sum_{\alpha=1,2}\psi_{\alpha}^{\dag}[\frac{p^{2}}{2m}-\mu]\psi_{\alpha}+V_{int} ,\notag\\
     &&V_{int}=\frac{1}{2}\int d^3\textbf{r}_1 d^3\textbf{r}_2\notag\\
     &&\!\!\sum_{\alpha\beta=1,2}\!\!\psi_{\alpha}^\dag(\textbf{r}_1)\psi_{\beta}^\dag(\textbf{r}_2)V_{\alpha\beta}(\textbf{r}_1-\textbf{r}_2)\psi_{\beta}(\textbf{r}_2)\psi_{\alpha}(\textbf{r}_1),\notag\\
\end{eqnarray}
where $V_{\alpha\beta}$ are interactions between atoms, and we assume $V_{\alpha\beta}=V_{\beta\alpha}$. $\psi_{1(2)}$ are two component bosonic field operators. $m$ is atomic mass and $\mu$ is chemical potential.
Here we generalize quantum field method  \cite{Abrikosov} to two-component Bosons.
When Bose-Einstein condensation occurs at zero-momentum in bosonic system, the field operators would take non-vanishing mean values, e.g.,
\begin{align}\label{eqn1}
   \langle\psi\rangle= \left(
 \begin{array}{cccc}   
   \langle \psi_1(\textbf{r}) \rangle \\  
   \langle \psi_2(\textbf{r}) \rangle\\  
\end{array}\right)= \left(
 \begin{array}{cccc}   
   \xi_{1} \\  
   \xi_{2}\\  
\end{array}\right)=\left(
 \begin{array}{cccc}   
   \sqrt{n_{1}} \\  
   \sqrt{n_{2}}\\  
\end{array}\right),
\end{align}
where $n_{\sigma=1,2}$ are condensate density for spin-up and spin-down component and $n_{1}+n_{2}=n_0$ gives the total condensate density.
The next step is taking the Bogoliubov substitution for field operators in Hamiltonian of eq. (\ref{hamiltonian}),
\begin{align}\label{substitution}
    \left(
 \begin{array}{cccc}   
 \psi_1(\textbf{r})\\  
   \psi_2(\textbf{r}) \\  
\end{array}\right)=\left(
 \begin{array}{cccc}   
  \xi_1+\phi_{1}(\textbf{r}) \\  
   \xi_2+\phi_{2}(\textbf{r})\\  
\end{array}\right),
\end{align}
where field operator $\phi_{1(2)}(r)$ has excluded the zero-momentum component.

After substitution with eq.(\ref{substitution}), the original Hamiltonian eq.(\ref{hamiltonian}) becomes
\begin{align}\label{eqn1}
&H\rightarrow K\equiv K_0+V_{int},\notag\\
&K_0=\int d^3\textbf{r}\sum_{\sigma=1,2}\phi^{\dag}_{\sigma}(\textbf{r})[\frac{p^2}{2}-\mu]\phi_{\sigma}(\textbf{r})-\mu[n_1+n_2],\notag\\
& V_{int}=E_0 +V_1+V_2+V_3+V_4+V_5+V_6+V_7,\notag\\
& E_0=\frac{1}{2}[n^{2}_{1}V_{11}(0)+2n_{1}n_2V_{12}(0)+n^{2}_{2}V_{22}(0)],\notag\\
& V_1=\frac{1}{2}\int d^3 \textbf{r}_1d^{3} \textbf{r}_2\sum_{\alpha \beta}\xi^{*}_\alpha \xi^{*}_\beta V_{\alpha\beta}(\textbf{r}_1,\textbf{r}_2)\phi_{\beta}(\textbf{r}_2)\phi_{\alpha}(\textbf{r}_1),\notag\\
&V_2=\frac{1}{2}\int d^3 \textbf{r}_1d^{3} \textbf{r}_2\sum_{\alpha \beta}\xi_\alpha \xi_\beta \phi^\dag_{\alpha}(\textbf{r}_1)\phi^\dag_{\beta}(\textbf{r}_2)V_{\alpha\beta}(\textbf{r}_1-\textbf{r}_2) ,\notag\\
&V_3=2(\frac{1}{2})\int d^3 \textbf{r}_1d^{3} \textbf{r}_2\sum_{\alpha\beta}\xi_\alpha\xi^{*}_{\beta}\phi^{\dag}_{\alpha}(\textbf{r}_1)V_{\alpha\beta}(\textbf{r}_1-\textbf{r}_2)\phi_{\beta}(\textbf{r}_2),\notag\\
&V_4=2(\frac{1}{2})\int d^3 \textbf{r}_1d^{3} \textbf{r}_2\sum_{\alpha\beta}\xi^{*}_{\beta}\xi_{\beta}\phi^{\dag}_{\alpha}(\textbf{r}_1)V_{\alpha\beta}(\textbf{r}_1-\textbf{r}_2)\phi_{\alpha}(\textbf{r}_1),\notag\\
&V_5=2(\frac{1}{2})\int d^3 \textbf{r}_1d^{3} \textbf{r}_2\sum_{\alpha\beta}\xi_{\beta}\phi^{\dag}_{\alpha}(\textbf{r}_1)\phi^{\dag}_{\beta}(\textbf{r}_2)V_{\alpha\beta}(\textbf{r}_1-\textbf{r}_2)\phi_{\alpha}(\textbf{r}_1),\notag\\
&V_6=2(\frac{1}{2})\int d^3 \textbf{r}_1d^{3} \textbf{r}_2\sum_{\alpha\beta}\xi^*_{\beta}\phi^{\dag}_{\alpha}(\textbf{r}_1)V_{\alpha\beta}(\textbf{r}_1-\textbf{r}_2)\phi_{\beta}(\textbf{r}_2)\phi_{\alpha}(\textbf{r}_2),\notag\\
&V_7=\frac{1}{2}\int d^3\textbf{r}_1d^3\textbf{r}_2\sum_{\alpha\beta}\!\!\phi_{\alpha}^\dag(\textbf{r}_1)\phi_{\beta}^\dag(\textbf{r}_2)V_{\alpha\beta}(\textbf{r}_1-\textbf{r}_2)\phi_{\beta}(\textbf{r}_2)\phi_{\alpha}(\textbf{r}_1),
\end{align}
where $d^3\textbf{r}\equiv dr_xdr_ydr_z$ and we set volume $V=1$, $m=1$.
If operator $\phi_{1(2)}(\textbf{r})$ acts on the non-interacting ground state, it gives zero. So Wick theorem for field operators can apply in two-component bosonic system. The figure.1 shows the various interaction processes in two-component boson condensate. The wavy line denotes the interaction between bosons $V_{\alpha\beta}(\textbf{r}_1-\textbf{r}_2)$, the  entering (leaving) dashed-lines represent the condensate factors $\xi^{*}_{\alpha}$ ($\xi_\alpha$).
\begin{figure}[h]
\begin{center}
\includegraphics[width=\columnwidth]{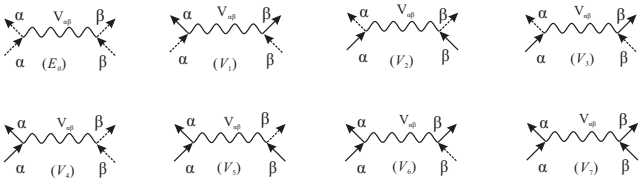}
\end{center}
\caption{(Color online) The interaction processes in two-component condensate. }
\label{figure1}
\end{figure}

Similar to usual single-component bosons \cite{Fetter}, we can define the normal Green's function
\begin{align}\label{normal}
  G_{\alpha \beta}(x,y)&\equiv-i\langle O|T[\phi_{\alpha}(x)\phi^{\dag}_{\beta}(y)]|O\rangle\notag\\
  &\equiv                
\{ \begin{array}{cccc}   
   -i\langle O|\phi_{\alpha}(x)\phi^{\dag}_{\beta}(y)|O\rangle,  & t_x>t_y \\  
  -i\langle O|\phi^\dag_{\beta}(y)\phi_{\alpha}(x)|O\rangle, & t_x<t_y
\end{array}\!\!\!
\end{align}
where $x\equiv\{\textbf{r}_x,t_x\}$ is space-time coordinate, $T$ is time ordering operator and $|O\rangle$ is exact ground state of $K$ in Heisenberg picture. The operator $A(t)=e^{iKt}A_Se^{-iKt}$ is also in Heisenberg picture and $A_S$ is operator in Schr\"{o}dinger picture.

The anomalous Green's functions are
\begin{align}\label{fanchang1}
F_{\alpha \beta}(x,y)&\equiv-i\langle O|T[\phi_{\alpha}(x)\phi_{\beta}(y)]|O\rangle\notag\\
&  \equiv \{                
 \begin{array}{cccc}   
   -i\langle O|\phi_{\alpha}(x)\phi_{\beta}(y)|O\rangle,  & t_x>t_y \\  
  -i\langle O|\phi_{\beta}(y)\phi_{\alpha}(x)|O\rangle,  & t_x<t_y
\end{array}\!\!\!
\end{align}
and \begin{align}\label{fanchang2}
F^+_{\alpha \beta}(x,y)&\equiv-i\langle O|T[\phi^{\dag}_{\alpha}(x)\phi^{\dag}_{\beta}(y)]|O\rangle\notag\\
&
  \equiv \{                
 \begin{array}{cccc}   
   -i\langle O|\phi^{\dag}_{\alpha}(x)\phi^{\dag}_{\beta}(y)|O\rangle,  & t_x>t_y \\  
  -i\langle O|\phi^{\dag}_{\beta}(y)\phi^{\dag}_{\alpha}(x)|O\rangle , & t_x<t_y.
\end{array}\!\!\!
\end{align}
From the definitions, we can see
\begin{align}\label{eqn1}
&F_{\alpha \beta}(x,y)=F_{\beta \alpha}(y,x),\notag\\
&F^{+}_{\alpha \beta}(x,y)=F^{+}_{\beta \alpha}(y,x).
\end{align}
The above Green's functions can be calculated in interaction picture, e.g.,
\begin{align}
&G_{\alpha \beta}(x,y)=-i\frac{\langle T[\tilde{\phi}_{\alpha}(x)\tilde{\phi}^{\dag}_{\beta}(y)S(-\infty,\infty)]\rangle}{\langle S(-\infty,\infty)\rangle},\notag\\
&F_{\alpha \beta}(p)=-i\frac{\langle T[\tilde{\phi}_{\alpha}(x)\tilde{\phi}_{\beta}(y)S(-\infty,\infty)]\rangle}{\langle S(-\infty,\infty)\rangle},\notag\\
&F^{+}_{\alpha \beta}(p)=-i\frac{\langle T[\tilde{\phi}^{\dag}_{\alpha}(x)\tilde{\phi}^{\dag}_{\beta}(y)S(-\infty,\infty)]\rangle}{\langle S(-\infty,\infty)\rangle},
\end{align}
where $\langle...\rangle$ denotes taking the average with respect to the non-interacting ground state. The operator $\tilde{A}(t)=e^{iK_0 t}A_{S}e^{-iK_0t}$ is operator in interaction picture and
\begin{align}
 &S(-\infty,\infty)\notag\\
 &=\sum_n \frac{(-i)^n}{n!}\int_{-\infty}^{\infty} dt_1dt_2...dt_n\langle T[\tilde{V}_{int}(t_1)\tilde{V}_{int}(t_2)...\tilde{V}_{int}(t_n)]\rangle,\notag\\
\end{align}
where $\tilde{V}_{int}(t)=e^{iK_0t}V_{int}e^{-iK_0t}$.

The above Green's functions can also be written as the perturbation series. The every order correction term can be similarly represented by Feynman diagram. The Feynman rules for perturbation series are similar to usual single component bosons except that there are more additional interaction lines  and vertices entailing spin indexes. The normal Green's function $G_{\alpha\beta}(x,y)$ can be represented by a thick solid line with two arrows of same directions (see figure.\ref{figure3}).
 The anomalous Green's functions $F$ or $F^+$  are represented with thick lines with two arrows with opposite directions (see figure.\ref{figure3}).  One should notes that the interactions do not flip spins (see figure.1). For every diagram, entering line number for spin-$\alpha$ is always equal to its leaving line number.
 The condensate factors $\xi^{*}_{\alpha}$ ($\xi_\alpha$) are represented by entering (leaving) dashed-lines.
In coordinate space, the Feynman rules for n-th order corrections of Green's function are listed here:

 (a):label each vertex with 4-coordinates $\{x_i\}$ and spin index $\{\alpha_i\}$;

(b): draw all the possible topological inequivalent connected diagrams (within such  diagrams, a diagram could not be obtained  from other diagrams by exchanging internal variables $x_i,\alpha_i \longleftrightarrow x_j,\alpha_j$) with n interaction lines;

(c): every wavy line represents the interaction potential $U_{\alpha\beta}(x_1-x_2)\equiv V_{\alpha\beta}(\textbf{r}_1-\textbf{r}_2)\delta(t_1-t_2)$;

(d): every thin solid line represents the Green's function $G^{0}_{\alpha\beta}(x,y)$ of non-interacting bosons;

(e): integral over all the internal space-time variables $\{x_i\}$;

(g): multiply a factor from condensate lines $(\xi^{*}_{1}\xi_{1})^{s_1}(\xi^{*}_{2}\xi_{2})^{s_2}$, where $s_{1(2)}$ is the leaving condensate line for spin-up(down);

(f): sum over all the internal spin-index $\{\alpha_i\}$;

(g): finally multiply a factor $i^{n-s_1-s_2}$.

\begin{figure}[h]
\begin{center}
\includegraphics[width=\columnwidth]{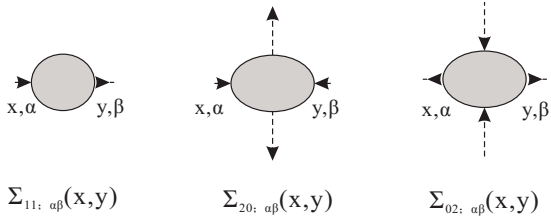}
\end{center}
\caption{(Color online) the self-energy diagram. }
\label{figure2}
\end{figure}

\begin{figure}[h]
\begin{center}
\includegraphics[width=\columnwidth]{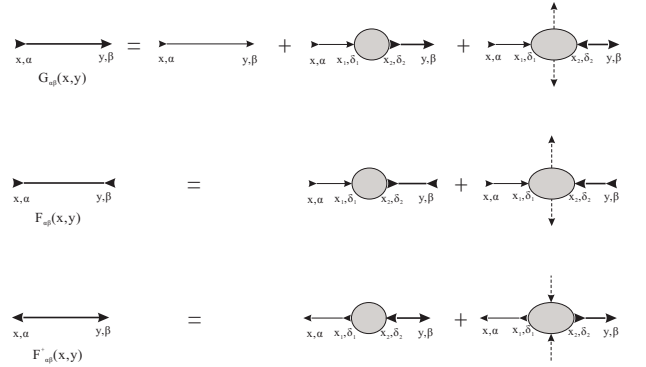}
\end{center}
\caption{(Color online) Dyson-Beliaev's equation in diagram from. }
\label{figure3}
\end{figure}

In bosonic system, apart from the normal self-energy $\Sigma_{11;\alpha \beta}(x,y)$,  there are two anomalous self-energies $\Sigma_{20;\alpha\beta}(x,y)$ and $\Sigma_{02;\alpha \beta}(x,y)$. Here the first number in the first pair of subscripts denotes  the number of entering non-condensate external lines, the second one  labels the number of leaving non-condensate lines \cite{Abrikosov}. While the second pair of subscripts denotes the spin-index for external lines.   So the Green's function $G, F,F^+$ and self energy $\Sigma_{11}, \Sigma_{20},\Sigma_{02}$ are all $2\times 2 $ matrices in spin space. The diagram representations for various self-energies are shown in figure. \ref{figure2}.

Taking spin-indexes into account, the Dyson-Beliaev's equation for Green's functions \cite{Beliaev,Abrikosov} can be written as
\begin{align}\label{eqn1}
&G_{\alpha \beta}(x,y)=G^{0}_{\alpha \beta}(x,y)\notag\\
&+\sum_{\delta_1 \delta_2}\int dx_1 dx_2[G^{0}_{\alpha \delta_1}(x,x_1)\Sigma_{11;\delta_{1} \delta_{2}}(x_1,x_2)G_{\delta_2 \beta}(x_2,y)\notag\\
&+G^{0}_{\alpha \delta_1}(x,x_1)\Sigma_{20;\delta_1 \delta_2}(x_1,x_2)F^+_{\delta_2 \beta}(x_2,y)],\notag\\
&F_{\alpha \beta}(x,y)\notag\\
&=\sum_{\delta_1\delta_2}\int dx_1dx_2[G^{0}_{\alpha \delta_1}(x,x_1)\Sigma_{11;\delta_1\delta_2}(x_1,x_2)F_{\delta_2\beta}(x_2,y)\notag\\
&+G^{0}_{\alpha \delta_1}(x,x_1)\Sigma_{20;\delta_1\delta_2}(x_1,x_2)G_{\beta \delta_2}(y,x_2)],\notag\\
&F^{+}_{\alpha\beta}(x,y)\notag\\
&=\sum_{\delta_1 \delta_2}\int dx_1dx_2[G^{0}_{\delta_1 \alpha}(x_1,x)\Sigma_{11;\delta_2 \delta_1}(x_2,x_1)F^{+}_{\delta_2 \beta}(x_2,y)\notag\\
&+G^{0}_{\delta_1\alpha}(x_1,x)\Sigma_{02;\delta_1\delta_2}(x_1,x_2)G_{\delta_2\beta}(x_2,y)],
\end{align}
where $dx\equiv dr_xdr_ydr_zdt_x$.
The figure.\ref{figure3} shows the Dyson-Beliaev's equation in diagram form.

For uniform system, the Green's functions are functions of coordinate difference, i.e.,
\begin{align}\label{eqn1}
&G_{\alpha \beta}(x,y)=G_{\alpha \beta}(x-y),\notag\\
&F_{\alpha \beta}(x,y)=F_{\alpha \beta}(x-y),\notag\\
&F^{+}_{\alpha \beta}(x,y)=F^{+}_{\alpha \beta}(x-y).
\end{align}
In the 4-momentum space of $p\equiv\{\textbf{q},\omega\}$, i.e.,
\begin{align}\label{eqn1}
&G_{\alpha \beta}(x,y)=\frac{1}{(2\pi)^4}\int dpG_{\alpha \beta}(p)e^{ip(x-y)},\notag\\
&F_{\alpha \beta}(x,y)=\frac{1}{(2\pi)^4}\int dpF_{\alpha \beta}(p)e^{ip(x-y)},\notag\\
&F^{+}_{\alpha \beta}(x,y)= \frac{1}{(2\pi)^4}\int dpF^{+}_{\alpha \beta}(p)e^{ip(x-y)},\notag\\
&U_{\alpha\beta}(x_1-x_2)= \frac{1}{(2\pi)^4}\int dpU_{\alpha\beta}(p)e^{ip(x-y)},\notag\\
&U_{\alpha\beta}(p)=\int dx U_{\alpha\beta}(x)=V_{\alpha\beta}(\textbf{q}),
\end{align}
where $dp\equiv d^3\textbf{q}dt$ and $px\equiv \textbf{q}\cdot \textbf{r}_x-\omega t_x$,
the Green's functions are
\begin{align}\label{greenfunction}
&G_{\alpha \beta}(p)\!\!=\!\!\sum_n[\frac{\langle O|\phi_{\alpha,\textbf{q}}|n\rangle\langle n|\phi^{\dag}_{\beta,\textbf{q}}|O\rangle}{\omega+i\eta-\omega_{n0}}\!-\!\frac{\langle O|\phi^{\dag}_{\beta,\textbf{q}}|n\rangle\langle n|\phi_{\alpha,\textbf{q}}|O\rangle}{\omega-i\eta+\omega_{n0}}],\notag\\
&F_{\alpha \beta}(p)=F_{\beta \alpha}(-p)\notag\\
&=\sum_n[\frac{\langle O|\phi_{\alpha,\textbf{q}}|n\rangle\langle n|\phi_{\beta,-\textbf{q}}|O\rangle}{\omega+i\eta-\omega_{n0}}-\frac{\langle O|\phi_{\beta,-\textbf{q}}|n\rangle\langle n|\phi_{\alpha,\textbf{q}}|O\rangle}{\omega-i\eta+\omega_{n0}}],\notag\\
&F^{+}_{\alpha \beta}(p)=F^{+}_{\beta \alpha}(-p)\notag\\
&=\sum_n[\frac{\langle O|\phi^{\dag}_{\alpha,-\textbf{q}}|n\rangle n|\phi^{\dag}_{\beta,\textbf{q}}|O\rangle}{\omega+i\eta-\omega_{n0}}-\frac{\langle O|\phi^{\dag}_{\beta,\textbf{q}}|n\rangle\langle n|\phi^{\dag}_{\alpha,-\textbf{q}}|O\rangle}{\omega-i\eta+\omega_{n0}}],
\end{align}
where excitation energy $\omega_{n0}\equiv E_n-E_0$. $E_n$, $|n\rangle$ are exact eigenenergy and eigenstates of $K$, respectively.
For uniform system, the Dyson-Beliaev's equation can be written in momentum space
\begin{align}\label{eqn1}
&G(p)=G^{0}(p)+G^{0}(p)\Sigma_{11}(p)G(p)+G^{0}(p)\Sigma_{20}(p)F^+(p),\notag\\
&F(p)=G^{0}(p)\Sigma_{11}(p)F(p)+G^{0}(p)\Sigma_{20}(p)G^{t}(-p),\notag\\
&F^{+}(p)=G^{0t}(-p)\Sigma_{11}^{t}(-p)F^{+}(p)+G^{0t}(-p)\Sigma_{02}(p)G(p),
\end{align}
where $A^{t}$ denote the matrix transpose  of $A$.
Furthermore, we can write the Dyson-Beliaev's equation in more compact way. Let's first define the column field operator
\begin{align}\label{eqn1}
\Psi(x)=\left(
\begin{array}{cccc}   
  \phi_{1}(x)\\  
   \phi_{2}(x)\\  
   \phi^{\dag}_{1}(x)\\
   \phi^{\dag}_{2}(x)\\
\end{array}\right).
\end{align}
The above Green's functions of Eqs.(\ref{normal}), (\ref{fanchang1}) and (\ref{fanchang2}) can be written as \cite{Fetter}
\begin{align}\label{eqn1}
\textbf{G}_{\alpha,\beta}(x,y)=-i\langle O|T[\Psi_{\alpha}(x)\Psi^{\dag}_{\beta}(y)]|O\rangle.
\end{align}
The Dyson-Beliaev's equation is taken the usual Dyson's equation form
\begin{align}\label{eqn1}
\textbf{G}(x,y)=\textbf{G}^{0}(x,y)+\textbf{G}^{0}(x,x_1)\Sigma(x_1,x_2)\textbf{G}(x_2,y),
\end{align}
and
\begin{align}\label{eqn1}
&\textbf{G}(x,y)=\left(
\begin{array}{cccc}   
   G(x,y)&F(x,y)\\
   F^{+}(x,y)&G^{t}(y,x)\\
\end{array}\right)_{4\times4},\notag\\
&\Sigma(x,y)=\left(
\begin{array}{cccc}   
   \Sigma_{11}(x,y)&\Sigma_{20}(x,y)\\
   \Sigma_{02}(x,y)&\Sigma^{t}_{11}(y,x)\\
\end{array}\right)_{4\times4}.
\end{align}
In uniform system, we apply a Fourier transformation to above equation; we get the Dyson-Beliaev's equation in 4-momentum space
\begin{align}\label{eqn1}
\textbf{G}(p)=\textbf{G}^{0}(p)+\textbf{G}^{0}(p)\Sigma(p)\textbf{G}(p),
\end{align}
where
\begin{align}
&\textbf{G}^{0}(p)=\textbf{diag}\notag\\
&\{\frac{1}{\omega+i\eta-\frac{q^2}{2}+\mu},\frac{1}{\omega+i\eta-\frac{q^2}{2}+\mu},\notag\\
&\frac{1}{-\omega+i\eta-\frac{q^2}{2}+\mu},\frac{1}{-\omega+i\eta-\frac{q^2}{2}+\mu}\} ,
\end{align}
and
\begin{align}\label{eqn1}
&\textbf{G}(p)=\left(
\begin{array}{cccc}   
   G(p)&F(p)\\
   F^{+}(p)&G^{t}(-p)\\
\end{array}\right)_{4\times4},\notag\\
&\Sigma(p)=\left(
\begin{array}{cccc}   
   \Sigma_{11}(p)&\Sigma_{20}(p)\\
   \Sigma_{02}(p)&\Sigma^{t}_{11}(-p)\\
\end{array}\right)_{4\times4}.
\end{align}
From the above equation, we get the self-energy in terms of Green's function
\begin{align}\label{selfenergy}
\Sigma(p)=(\textbf{G}^{0}(p))^{-1}-\textbf{G}^{-1}(p).
\end{align}
From the above eqs. (\ref{selfenergy}) and eq.(\ref{greenfunction}), the anomalous self-energy satisfy
\begin{align}\label{selfenergy1}
\Sigma_{20;\beta\alpha}(-p)=\Sigma_{20; \alpha \beta}(p),\notag\\
\Sigma_{02;\beta\alpha}(-p)=\Sigma_{02; \alpha \beta}(p).
\end{align}
 In momentum space, the Feynman rules for n-th order corrections of Green's function are:

(a): draw all the possible topological inequivalent  connected diagrams with n interaction lines;

(b): label each vertex with spin index $\{\alpha_i\}$; assign a direction for every line; at every vertex, the momenta are conserved;

(c): every directed thin solid line represents the Green's function $G^{0}_{\alpha\beta}(p)=\frac{\delta_{\alpha\beta}}{\omega+i\eta-q^2/2+\mu}$ of non-interacting bosons;

(d): every wavy line represents interaction potential $U_{\alpha\beta}(p)\equiv V_{\alpha\beta}(\textbf{q})$;

(e): integral all the momenta which could not be determined by momentum conservation; for every momentum integral, multiply a factor $\frac{1}{(2\pi)^4}$;

(f): multiply a factor from condensate lines $(\xi^{*}_{1}\xi_{1})^{s_1}(\xi^{*}_{2}\xi_{2})^{s_2}$, where $s_{1(2)}$ is the leaving condensate lines for spin-up(down);

(g): sum over all the internal spin-indexes $\{\alpha_i\}$;

(h): finally multiply a factor $i^{n-s_1-s_2}$.

Because the Hamiltonian $K$ is real in the Fock space of momentum, the eigen-states of $K$, i.e., $|n\rangle$ are all real.
So the matrix elements of $a_{\textbf{q}}$, $a^{\dag}_{\textbf{q}}$ between arbitrary two eigen-states $\langle m|a_\textbf{q}|n\rangle$ or $\langle m|a^{\dag}_\textbf{q}|n\rangle$ can be taken as real \cite{Lifshitz}.
From eq. (\ref{greenfunction}), we get
\begin{align}\label{selfenergy2}
&G_{\alpha \beta}(p)=G_{\beta \alpha}(p),\notag\\
&F^{+}_{\alpha \beta}(-p)=F_{\alpha \beta }(p).
\end{align}
At zero-frequency of $\omega\pm i\eta=0$, all the above Green's functions  of eqs.(\ref{greenfunction}) in momentum space are real.
Using equation (\ref{selfenergy}) and (\ref{selfenergy2}), we furthermore get that the self-energy satisfy
\begin{align}\label{selfenergy3}
& \Sigma_{11;\alpha \beta}(p)=\Sigma_{11;\beta \alpha }(p),\notag\\
&\Sigma_{02;\alpha \beta}(-p)=\Sigma_{20,\alpha \beta }(p).
\end{align}

The Green's function  can be expressed in term of self-energy
\begin{align}\label{Greenfunction1}
 \textbf{G}(p)=(\textbf{I}-\textbf{G}^{0}(p)\Sigma(p))^{-1}\textbf{G}^{0}(p).
 \end{align}
Using the properties of self-energy eqs. (\ref{selfenergy1}) and (\ref{selfenergy3}), the self-energy matrix can be written as
\begin{align}
\Sigma(p)=\left(
\begin{array}{cccc}   
  \Sigma_{11;11}(p)&\Sigma_{11;12}(p)&\Sigma_{20;11}(p)&\Sigma_{20;12}(p)\\  
   \Sigma_{11;12}(p)&\Sigma_{11;22}(p)&\Sigma_{20;12}(-p)&\Sigma_{20;22}(p)\\  
  \Sigma_{20;11}(p)&\Sigma_{20;12}(-p)&\Sigma_{11;11}(-p)&\Sigma_{11;12}(-p)\\
   \Sigma_{20;12}(p)&\Sigma_{20;22}(p)&\Sigma_{11;12}(-p)&\Sigma_{11;22}(-p)\\
\end{array}\right).
\end{align}

 The ground state energy (grand potential) is
 \begin{align}
 \Omega(\mu,n_1,n_2)=\langle O| K |O\rangle.
 \end{align}
 Here there are three parameters, $\mu$, $n_1$ and $n_2$. They are determined by particle number equation
\begin{align}
n=n_0+\sum_{\alpha}\int d^{3}\textbf{r} i G_{\alpha\alpha}(\textbf{r}t,\texttt{r}t+0_+),
 \end{align}
 and  minimized conditions of ground state energy with respect to $n_1$ and $n_2$, i.e.,
\begin{align}
&\frac{\partial \Omega }{\partial n_1}=0, \Rightarrow \mu=\langle\frac{\partial V_{int}}{\partial n_1}\rangle,\notag\\
&\frac{\partial \Omega }{\partial n_2}=0, \Rightarrow \mu=\langle\frac{\partial V_{int}}{\partial n_2}\rangle.
\label{mu}
 \end{align}

\section{generalized Hugenholtz-Pines relation }
Next let's show a generalized Hugenholtz-Pines relation hold for two component bosons. Similar to single-component bosons, a crucial point is to note that the self-energy diagrams at zero momentum-frequency can be obtained from diagrams of interaction energy through replacing two condensate lines  with their corresponding non-condensate lines (with same spin-indexes and $p=0$) \cite{Pines1959,Abrikosov}. $\langle V_{int}\rangle _{m_1,m_2,m_3; \{s_1,\tilde{s}_1\},\{s_2,\tilde{s}_2\}}$  denotes the $m-th$ order corrections to interaction energy  from diagrams with $m_1+m_2+m_3=m+1$ interaction lines (because  diagrams which contribute m-th corrections for interaction energy have m+1 interaction lines in total), e.g., $m_1$ $V_{11}$ lines, $m_2$ $V_{12}$ lines, $m_3$ $V_{22}$ lines, $s_{1(2)}$ entering condensate line for spin-up (down) and $\tilde{s}_{1(2)}$ leaving condensate line for spin-up (down).
 Because the interactions do not flip the spins, the number of  entering condensate lines $s_{1(2)}$ should be equal to that of leaving condensate lines $\tilde{s}_{1(2)}$, i.e.,
\begin{align}
&\tilde{s}_1=s_1,\notag\\
&\tilde{s}_2=s_2.
 \end{align}

So there are a $(\xi^{*}_{1}\xi_{1})^{s_1}(\xi^{*}_{2}\xi_{2})^{s_2}=n_{1}^{s_1}n_{2}^{s_2}$ factor in such diagram of $\langle V_{int}\rangle _{m_1,m_2,m_3; \{s_1,\tilde{s}_1\},\{s_2,\tilde{s}_2\}}$ .  From Eq.(\ref{mu}), the chemical potential can be written as
\begin{align}
&\mu=\sum_{m_1,m_2,m_3,s_1,s_2}\frac{s_1}{n_1}\langle V_{int}\rangle_{m_1,m_2,m_2;\{s_1,s_1\},\{s_2,s_2\}},\notag\\
&\mu=\sum_{m_1,m_2,m_3,s_1,s_2}\frac{s_2}{n_2}\langle V_{int}\rangle_{m_1,m_2,m_2;\{s_1,s_1\},\{s_2,s_2\}}.
\label{mu1}
 \end{align}

 In addition, the diagrams which have $m+1$-th order corrections ($(\Sigma_{11,11}(\textbf{0},0))_{m+1;m_1,m_2,m_3; \{s_1-1,s_1-1\},\{s_2,s_2\}}$) to normal self-energy can be obtained from diagrams of $\langle V_{int}\rangle_{m_1,m_2,m_2;\{s_1,s_1\},\{s_2,s_2\}}$ by replacing a entering condensate line $\xi^{*}_{1}$ and a leaving condensate line $\xi_{1}$ of spin-up (see figure.2). There are $s_1\times s_1$ ways to replace. For every diagram of $\langle V_{int}\rangle_{m_1,m_2,m_2;\{s_1,s_1\},\{s_2,s_2\}}$, there are $s_1\times s_1$ corresponding diagrams to contribute to $(\Sigma_{11,11}(\textbf{0},0))_{m+1;m_1,m_2,m_3; \{s_1,s_1\},\{s_2,s_2\}}$ . So the self-energy $\Sigma_{11;11}(\textbf{0},0)$ can be written as
\begin{align}\label{selfenergy11}
&\Sigma_{11;11}(\textbf{0},0)\notag\\
&=\sum_{m_1,m_2,m_3,s_1,s_2}(\Sigma_{11,11}(\textbf{0},0))_{m+1;m_1,m_2,m_3;\{s_1-1,s_1-1\},\{s_2,s_2\}}\notag\\
&=\sum_{m_1,m_2,m_3,s_1,s_2}\frac{s^{2}_{1}}{n_1}\langle V_{int}\rangle_{m_1,m_2,m_2;\{s_1,s_1\},\{s_2,s_2\}}.
 \end{align}

The anomalous self-energy corrections from $m+1$-th order perturbation $(\Sigma_{20,11}(\textbf{0},0))_{m+1;m_1,m_2,m_3;\{s_1-2,s_1\},\{s_2,s_2\}}$  can be obtained from $\langle V_{int}\rangle_{m_1,m_2,m_2;\{s_1,s_1\},\{s_2,s_2\}}$ by
replacing two entering condensate lines $\xi^{*}_{1}$ of spin-up (see figure.2). For every diagram of $\langle V_{int}\rangle_{m_1,m_2,m_2;\{s_1,s_1\},\{s_2,s_2\}}$, there are $s_1\times (s_1-1)$ corresponding diagrams to contribute to $(\Sigma_{20,11}(\textbf{0},0))_{m+1;m_1,m_2,m_3;\{s_1-2,s_1\},\{s_2,s_2\}}$ , so the self-energy $\Sigma_{20;11}(\textbf{0},0)$ is
\begin{align}\label{selfenergy22}
&\Sigma_{20;11}(\textbf{0},0)\notag\\
&=\sum_{m_1,m_2,m_3,s_1,s_2}(\Sigma_{20,11}(\textbf{0},0))_{m+1;m_1,m_2,m_3;\{s_1-2,s_1\},\{s_2,s_2\}}\notag\\
&=\sum_{m_1,m_2,m_3,s_1,s_2}\frac{s^{2}_{1}-s_1}{n_1}\langle V_{int}\rangle_{m_1,m_2,m_2;\{s_1,s_1\},\{s_2,s_2\}}.
 \end{align}
Comparing eq. (\ref{selfenergy11}), (\ref{selfenergy22}) with (\ref{mu1}), we get \cite{Abrikosov}
\begin{align}
\mu=\Sigma_{11,11}(\textbf{0},0)-\Sigma_{20,11}(\textbf{0},0).
 \end{align}

Similarly, we also get
\begin{align}
\mu=\Sigma_{11,22}(\textbf{0},0)-\Sigma_{20,22}(\textbf{0},0).
 \end{align}

 As for off-diagonal self-energy $\Sigma_{11,12}(\textbf{0},0)$, we can get the $m+1$-th order corrections of diagrams $\Sigma_{11,12}(\textbf{0},0)_{m+1;m_1,m_2,m_3;\{s_1-1,s_1\};\{s_2,s_2-1 \}}$ by replacing one entering condensate line $\xi^{*}_{1}$ and one leaving condensate line $\xi_{2}$ from diagrams of $\langle V_{int}\rangle_{m_1,m_2,m_2;\{s_1,s_1\},\{s_2,s_2\}}$. For every diagram of $\langle V_{int}\rangle_{m_1,m_2,m_2;\{s_1,s_1\},\{s_2,s_2\}}$,  there are $s_1 s_2$ corresponding diagrams which have contributions to self-energy, so the self-energy $\Sigma_{11;12}(\textbf{0},0)$
\begin{align}\label{nondiagnal11}
&\Sigma_{11;12}(\textbf{0},0)\notag\\
&=\sum_{m_1,m_2,m_3,s_1,s_2}(\Sigma_{11,12}(\textbf{0},0))_{m+1;\{s_1-1,s_1\},\{s_2,s_2-1\}}\notag\\
&=\sum_{m_1,m_2,m_3,s_1,s_2}\frac{s_{1} s_2}{\sqrt{n_1n_2}}\langle V_{int}\rangle_{m_1,m_2,m_2;\{s_1,s_1\},\{s_2,s_2\}}.
 \end{align}
 Based on the similar reasons, we have
\begin{align}\label{nondiagnal12}
&\Sigma_{20;12}(\textbf{0},0)\notag\\
&=\sum_{m_1,m_2,m_3,s_1,s_2}\frac{s_{1} s_2}{\sqrt{n_1n_2}}\langle V_{int}\rangle_{m_1,m_2,m_2;\{s_1,s_1\},\{s_2,s_2\}}.
 \end{align}
 From the above eq.(\ref{nondiagnal11}) and (\ref{nondiagnal12}), we have
 \begin{align}
&\Sigma_{11;12}(\textbf{0},0)-\Sigma_{20,12}(\textbf{0},0)=0.
 \end{align}
Similarly,
\begin{align}
&\Sigma_{11;21}(\textbf{0},0)-\Sigma_{20,21}(\textbf{0},0)=0.
 \end{align}
 So the generalized Hugenholtz-Pines relation
 \begin{align}\label{pinesrelation}
&&\!\!\left( \!\!\!                
 \begin{array}{cccc}   
   \mu  & 0 \\  
   0& \mu\\  
\end{array}\!\!\!\right)=\!\!\left( \!\!\!                
 \begin{array}{cccc}   
   \Sigma_{11,11}(\textbf{0},0)  & \Sigma_{11,12}(\textbf{0},0) \\  
   \Sigma_{11,21}(\textbf{0},0)& \Sigma_{11,22}(\textbf{0},0)\\  
\end{array}\!\!\!\right)\notag\\
&&-\!\!\left( \!\!\!                
 \begin{array}{cccc}   
   \Sigma_{20,11}(\textbf{0},0) & \Sigma_{20,12}(\textbf{0},0) \\  
   \Sigma_{20,21}(\textbf{0},0)& \Sigma_{20,22}(\textbf{0},0)\\  
\end{array}\!\!\!\right)
\end{align}
 hold.
 Similarly, we also have
 \begin{align}\label{pinesrelation1}
&&\!\!\left( \!\!\!                
 \begin{array}{cccc}   
   \mu  & 0 \\  
   0& \mu\\  
\end{array}\!\!\!\right)=\!\!\left( \!\!\!                
 \begin{array}{cccc}   
   \Sigma_{11,11}(\textbf{0},0)  & \Sigma_{11,12}(\textbf{0},0) \\  
   \Sigma_{11,21}(\textbf{0},0)& \Sigma_{11,22}(\textbf{0},0)\\  
\end{array}\!\!\!\right)\notag\\
&&-\!\!\left( \!\!\!                
 \begin{array}{cccc}   
   \Sigma_{02,11}(\textbf{0},0) & \Sigma_{02,12}(\textbf{0},0) \\  
   \Sigma_{02,21}(\textbf{0},0)& \Sigma_{02,22}(\textbf{0},0)\\  
\end{array}\!\!\!\right).
\end{align}
 The generalized Hugenholtz-Pines relation has also been discussed in polar phase for spin-1 Bose-Einstein condensate \cite{Szirmai} and for explicitly broken SO(N) (SU(N)) symmetry case \cite{Watabe}.

\section{Two branch phonon excitations }
Note that the Hamiltonian is invariant with respect to transformation
\begin{align}
a_\textbf{k}\rightarrow a_{-\textbf{k}},a^{\dag}_{\textbf{k}}\rightarrow a^{\dag}_{-\textbf{k}},
\end{align}
so the Green's functions and self-energies are also invariant under the transformation $\textbf{k}\rightarrow-\textbf{k}$ \cite{Bogoliubov}. Taking  the isotropy of system  into account, so the Green's functions and self-energies should depend on the $\textbf{q}$ only through $q^2$.
In the following, we assume the self-energy $\Sigma(\textbf{q},\omega)$ can be expanded as series in terms of $\omega$ and $q^2$ near $\Sigma(\textbf{0},0)$ \cite{Lifshitz}, i.e.,
 \begin{align}
& \Sigma_{11,11}(p)=\Sigma_{11,11}(\textbf{0},0)+a_{111}\omega+a_{112}\omega^2+a_{113}q^2,\notag\\
&\Sigma_{11,12}(p)=\Sigma_{11,12}(\textbf{0},0)+a_{121}\omega+a_{122}\omega^2+a_{123}q^2,\notag\\
& \Sigma_{11,22}(p)=\Sigma_{11,22}(\textbf{0},0)+a_{221}\omega+a_{222}\omega^2+a_{223}q^2,\notag\\
& \Sigma_{20,11}(p)=\Sigma_{11,11}(\textbf{0},0)+b_{112}\omega^2+b_{113}q^2,\notag\\
&\Sigma_{20,12}(p)=\Sigma_{20,12}(\textbf{0},0)+b_{121}\omega+b_{122}\omega^2+b_{123}q^2,\notag\\
& \Sigma_{20,22}(p)=\Sigma_{20,22}(\textbf{0},0)+b_{222}\omega^2+b_{223}q^2.
\label{selfenergyexpansion}
\end{align}
 Due to $\Sigma_{20,\alpha \alpha}(p)=\Sigma_{20,\alpha \alpha}(-p)$ (see eq.(\ref{selfenergy1})), the diagonal elements of anomalous self-energy have no linear terms in $\omega$.
 Substituting the above expansion in the Green's function eq.(\ref{Greenfunction1}) and using the Hugenholtz-Pines relation eq.(\ref{pinesrelation}), from the poles of Green's function, i.e.,
 \begin{align}\label{Det}
Det[\textbf{I}-\textbf{G}^{0}(\textbf{q},\omega)\Sigma(\textbf{q},\omega)]=0,
\end{align}
 we generally get two branch phonon excitations with linear dispersion, i.e.,
  \begin{align}
\omega_1(q)=s_1q, \notag\\
\omega_2(q)=s_2q,
\end{align}
where $s_{1(2)}$ are two sound velocities.
At the zero-frequency of $\omega\pm i\eta=0$, from the eq.(\ref{Greenfunction1}), we find the leading terms of Green's functions satisfy
\begin{align}
F_{\alpha \beta}(\textbf{q},0)=-G_{\alpha \beta}(\textbf{q},0)\propto 1/q^2.
\end{align}
The above result is very similar to single-component case where $F(\textbf{q},0)=-G(\textbf{q},0)$ as $q\rightarrow0$.
At the zero-frequency, we note that here the time-ordering Green's functions are exactly same to the retarded Green's functions appeared in main text.

 \section{existence of quadratic dispersion excitation for U(2) invariant interactions}
When interaction strength are all equal $V_{11}=V_{12}=V_{22}$, the whole system has U(2) symmetry. The order parameter  for BEC has U(2) degeneracy, we can take
\begin{eqnarray}\label{eqn1}
    \langle\psi\rangle=\frac{\sqrt{n_0}}{\sqrt{2}}\left(
 \begin{array}{cccc}   
   1 \\  
   1\\  
\end{array}\right),
\end{eqnarray}
for example without loss generality. Within Bogoliubov method, one can get two branch gapless excitations, one is linear phonon $\omega_1(q)\sim q$ as $q\rightarrow0$, the other is quadratic dispersion $\omega_2(q)\sim q^2$ \cite{Goldstein,Pitaevskii}.
In the following, we will show the existence of above two types of excitations up to arbitrary order of perturbation theory.

  Let's first prove that the diagonal elements and off-diagonal elements of  anomalous self-energy are equal at $p=\{\textbf{q},\omega\}=0$, i.e.,
\begin{align}
\Sigma_{20;11}(\textbf{0},0)=\Sigma_{20;12}(\textbf{0},0)=\Sigma_{20;22}(\textbf{0},0).
 \end{align}
From the diagrams of $\langle V_{int}\rangle_{m_1,m_2,m_3;\{s_1,s_1\},\{s_2,s_2\}}$ which have m-th order corrections to interaction energy,  we replace two entering condensate lines of spin-up, i.e., $\xi^{*}_1$, we get diagrams of  $(\Sigma_{20;11}(\textbf{0},0))_{m+1; m_1,m_2,m_3,\{s_1-2,s_1\},\{s_2,s_2\}}$. The contribution of $\Sigma_{20;12}(\textbf{0},0)_{m+1; m_1,m_2,m_3,\{s_1-1,s_1\},\{s_2-1,s_2\}}$ can be obtained from $\langle V_{int}\rangle_{m_1,m_2,m_3;\{s_1,s_1\},\{s_2,s_2\}}$ by replacing one entering condensate line of spin-up ($\xi^{*}_1$) and a entering line of spin-down ($\xi^{*}_2$)(see figure.2).

Let's focus on a class of diagrams with $s_1+s_2=s$ pair  condensate lines in total. First of all, for very diagram in which all the condensate lines are spin-up, e.g., labeling them as $\langle V_{int}\rangle_{m,\{s,s\},\{0,0\}}$, there are $s*(s-1)$ corresponding diagrams to contribute to $\Sigma_{20;11}(\textbf{0},0)$, 0 diagram to contribute $\Sigma_{20;12}(\textbf{0},0)$. Next let's move to the diagrams which have $s-1$ pair condensate lines of spin-up and $1$ pair condensate line of spin-down, i.e., diagrams of $\langle V_{int}\rangle_{m,\{s-1,s-1\},\{1,1\}}$. However, for every diagram of $\langle V_{int}\rangle_{m,\{s,s\},\{0,0\}}$, there are $C_{s}^{1}$ corresponding diagrams of $\langle V_{int}\rangle_{m,\{s-1,s-1\},\{1,1\}}$. So for every diagram of $\langle V_{int}\rangle_{m,\{s,s\},\{0,0\}}$, there are  $C_{s}^{1} (s-1)*(s-2)$ diagrams in total to contribute to $\Sigma_{20;11}(\textbf{0},0)$; $C_{s}^{1}(s-1)*1$ diagrams have contributions in $\Sigma_{20;12}(\textbf{0},0)$. Similarly going on decreasing the pair number of condensate lines of spin-up, till all $s$ pair condensate lines are for spin-down component.  Then we get the total number of  the diagrams to contribute $\Sigma_{20;11}(\textbf{0},0)$ and $\Sigma_{20;12}(\textbf{0},0)$ respectively
\begin{eqnarray}\label{eqn1}
   &&N_1=\sum_{m=0}^{s}C_{s}^{m}(s-m)(s-m-1)=s(s-1)2^{s-2},\notag\\
   &&N_2=\sum_{m=0}^{s}C_{s}^{m}(s-m)(m)=s(s-1)2^{s-2},
\end{eqnarray}
and they are equal. Due to the equivalence between spin-up and spin-down components and equal interactions,  every diagram has same contribution in their respective self-energies. So we conclude
\begin{eqnarray}\label{eqn1}
   \Sigma_{20;11}(\textbf{0},0)=\Sigma_{20;12}(\textbf{0},0)=\Sigma_{20;22}(\textbf{0},0).
\end{eqnarray}

For U(2) invariant interaction bosons, due to the equivalence between spin-up and spin-down components, diagonal matrix elements of self-energies are equal and self-energies are symmetric matrices, i.e.,
\begin{align}
\Sigma_{11;11 }(p)=\Sigma_{11;22}(p),\notag\\
   \Sigma_{11;12}(p)=\Sigma_{11;21}(p),\notag\\
   \Sigma_{20;11 }(p)=\Sigma_{20;22}(p),\notag\\
   \Sigma_{20;12}(p)=\Sigma_{20;21}(p).
\end{align}
Consequently  the linear terms in $\omega$ are vanishing in the expansion of the anomalous self-energy near $p=\{\textbf{0},0\}$.
In additions, due to the above symmetric properties of self-energies, the Green's functions also have symmetric properties, i.e.,
\begin{align}
G_{11}(p)=G_{22}(p),\notag\\
   G_{12}(p)=G_{21}(p),\notag\\
   F_{11 }(p)=F_{22}(p),\notag\\
   F_{12}(p)=F_{21}(p),\notag\\
   F^{+}_{11 }(p)=F^{+}_{22}(p),\notag\\
   F^{+}_{12}(p)=F^{+}_{21}(p).
\end{align}

From the eq.(\ref{Det}), we get two equations for poles of Green's function
\begin{align}\label{eqn2}
&[-\omega+q^2/2-\mu+\Sigma_{11;11}(p)+\Sigma_{11;12}(p)]\notag\\
&[\omega+q^2/2-\mu+\Sigma_{11;11}(-p)+\Sigma_{11;12}(-p)]\notag\\
&=[\Sigma_{20;11}(p)+\Sigma_{20;12}(p)]^2,\notag\\
&[-\omega+q^2/2-\mu+\Sigma_{11;11}(p)-\Sigma_{11;12}(p)]\notag\\
&[\omega+q^2/2-\mu+\Sigma_{11;11}(-p)-\Sigma_{11;12}(-p)]\notag\\
&=[\Sigma_{20;11}(p)-\Sigma_{20;12}(p)]^2.
\end{align}
Further using the generalized Hugenholtz-Pines relation eq.(\ref{pinesrelation}), expansion of self-energies eq.(\ref{selfenergyexpansion}), eq. (\ref{eqn1}) and the absence of linear terms  of $\omega$ in the expansion of the anomalous self-energy ($\Sigma_{20;11(12)}$), the first equation gives a phonon $\omega_{1}(q)\sim q$; the second equation gives $\omega_{2}(q)\sim q^2$ . So there are always the quadratic dispersion excitation in $U(2)$ invariant interaction bosons.
As $q\rightarrow0$, we find the Green's functions satisfy
\begin{align}
&F_{11}(\textbf{q},0)+F_{12}(\textbf{q},0)=-[G_{11}(\textbf{q},0)+G_{12}(\textbf{q},0)]\propto 1/q^2.
\end{align}

\appendix*

\end{document}